\RequirePackage{fix-cm}
\documentclass[twocolumn]{svjour3}          
\smartqed  
\usepackage{graphicx}
\usepackage{multirow}
\usepackage{amsmath}
\usepackage{amsfonts}
\usepackage{cuted}
\usepackage{comment}
\usepackage{xcolor}
\usepackage{booktabs}
\usepackage{url}

\journalname{IJCV}
\begin{document}

\title{An Efficient Recurrent Adversarial Framework for Unsupervised Real-Time Video Enhancement

}


\author{Dario Fuoli$^1$ \and
        Zhiwu Huang$^{1,2}$ \and
        Danda Pani Paudel$^1$ \and 
        Luc Van Gool$^{1,3}$ \and
        Radu Timofte$^{1,4}$
}


\institute{$^1$Computer Vision Lab, ETH Z\"urich, Switzerland \\
              $^2$SAVG, Singapore Management University, Singapore\\
              $^3$VISICS, KU Leuven, Belgium\\
              $^4$CAIDAS, University of W\"urzburg, Germany\\
             \email{\{dario.fuoli, zhiwu.huang, paudel, vangool\}@vision.ee.ethz.ch; radu.timofte@uni-wuerzburg.de}
}

\date{ }

\maketitle

\begin{abstract}
Video enhancement is a challenging problem, more than that of stills, mainly due to high computational cost, larger data volumes and the difficulty of achieving consistency in the spatio-temporal domain. In practice, these challenges are often coupled with the lack of example pairs, which inhibits the application of supervised learning strategies. To address these challenges, we propose an efficient adversarial video enhancement framework that learns directly from unpaired video examples. In particular, our framework introduces new recurrent cells that consist of interleaved local and global modules for implicit integration of spatial and temporal information. The proposed design allows our recurrent cells to efficiently propagate spatio-temporal information across frames and reduces the need for high complexity networks. 
Our setting enables learning from unpaired videos in a cyclic adversarial manner, where the proposed recurrent units are employed in all architectures. Efficient training is accomplished by introducing one single discriminator that learns the joint distribution of source and target domain simultaneously.
The enhancement results demonstrate clear superiority of the proposed video enhancer over the state-of-the-art methods, in all terms of visual quality, quantitative metrics, and inference speed. Notably, our video enhancer is capable of enhancing over 35 frames per second of FullHD video (1080x1920). 
\end{abstract}

\section{Introduction}
Images or videos acquired from cameras are the outcome of algorithmic transformations of the sensor measurements. Such transformations are often modeled manually, while aiming to maximize the visual perceptual quality. 
Better sensor measurements obviously result in  higher perceptual quality. However, improving the sensor's output quality is not always an option due to several limitations, including but not limited to economical and/or physical constraints.
Moreover, images/videos of higher perceptual quality could also be carefully selected from a wide variety of data, or even enhanced manually to serve the same.

The perceptual quality of mediocre images/videos, e.g. acquired from consumer-grade cameras, can be improved by algorithmic enhancement techniques.
For this purpose, learning example-based enhancement is of very high interest in the literature, which is also found to be promising in \cite{ignatov2017dslr}. Example-based enhancement methods assume that the source (mediocre) and target (high quality) data examples are provided in some forms. The assumed forms however, either violate the practicality or limit the learning capabilities. We are concerned with the impractical assumption of the availability of pixel and frame-wise pairing of source and target video examples.

In the literature, the problem of frame-wise alignment between example video pairs, i.e. the synchronization problem, is circumvented by merely addressing the enhancement 
frame-by-frame, which essentially boils down to the unpaired image enhancement problem \cite{chen2018deep,ignatov2018wespe}. Here, the unpairdness refers to the unknown pixel-wise alignment (or the correspondences) between source and target images. The unpaired image enhancement methods are both tailored and suitable for image enhancement. When they are applied independently to video frames, the output videos are very likely to be temporally inconsistent. We refer temporal inconsistency to the inhomogeneous enhancement across video frames. Such inconsistency results in unnatural videos, thus of lower perceptual quality.  

Besides the spatio-temporal inconsistency, the problem of video enhancement invites new challenges of higher computational cost and larger data volumes, when compared to that of stills. Furthermore, video enhancement algorithms are often required to work in real-time, subject to hard time constraints in applications such as live TV, streaming services, and video acquisition using smartphones. This work addresses all the aforementioned challenges. Namely, (i) Learning from unpaired videos; (ii) Spatio-temporally consistent enhancement; (iii) Speed- and memory-efficient training; and (iv) Real-time inference. 
To tackle these challenges, we propose an efficient recurrent adversarial framework with the following listed contributions.

\noindent\textbf{Learning from unpaired videos:}
Inspired by \cite{chen2018deep,ignatov2018wespe}, we propose a new cyclic Generative Adversarial Network (GAN) based framework to learn the distribution map from source domain (input videos) to target domain (output videos) with desirable perceptual qualities, such that the pairdness of video examples is not required. The framework consists of two generators and a single discriminator. One of the two generators is trained to learn the quality map from source domain to target domain, while the other learns the reverse map (also referred as degrader). The discriminator is tasked to process both source and target videos. In particular, the discriminator takes a pair of videos -- one from source and the other from target -- as input, and learns the joint distribution between them. This facilitates us to make use of a single discriminator for both domains, leading to a simpler and more effective design.

\noindent\textbf{Spatio-temporal consistency:} We exploit a new recurrent design for both the generators and the discriminator, such that the desired consistency is directly implied.
In this regard, we introduce new recurrent cells that comprise interleaved local and global modules for better aggregation of spatial and temporal information. This design enables a fully connected view from the past spatio-temporal information in a sequence to each pixel-location in the output.

\noindent\textbf{Efficient training:}  Our generators and discriminator efficiently process videos using an improved recurrent latent space propagation \cite{rlsp2019}. Such a design not only efficiently propagates the spatio-temporal information, but also supports us to build smaller networks. This greatly decreases the complexity of all the involved discriminator and generators, thereby making our network both memory- and computation-efficient during training. Note that efficient training is necessary for high resolution videos due to the large data volumes.

\noindent\textbf{Real-time inference:} The real-time performance of our network is a direct consequence of our efficient design. This specifically applies to the generator that translates mediocre videos into high quality ones; in other words - the enhancer. 
Up to our knowledge, the proposed method is the first to achieve real-time performance for high definition deep video enhancement tasks.
The attained inference speed is over 35 frames per second (fps), which we regard to be suitable for many practical applications.

In order to achieve state-of-the-art results in real time, along with the necessary training efficiency and stability towards a more elegant data-driven framework, the following considerations were made in this paper. 
The desired properties are achieved by avoiding: (i) data/domain specific handcrafted losses and division into auxiliary subtasks \footnote{Total variation loss, identity loss, Y/CbCr separation, etc.}; (ii) normalization in recurrent adversarial setting, which inefficiently reduces the discriminator's capacity; (iii) and two separate discriminators -- each for source and target domain. To achieve better training stability, a temporal padding strategy is introduced. 
This enables the network to accumulate information in the latent state, without exposing the unavoidable initialization to the discriminator. Such exploitation was observed to be very effective in our experiments, with regard to training stability.

To evaluate the proposed recurrent framework, we rely on the Vid3oC dataset \cite{kim2019vid3oc} as well as a new large-scale dataset that includes much more videos. Both datasets use target videos as those captured by better sensors, when compared to that of the source.
This dataset consideration allows us to collect data in a very practical manner, e.g. by using a rig where both cameras of source and target are mounted side-by-side.
Although this offers the possibility of collecting data with the same video content, their pixel and frame-wise alignment is not possible due to differences in viewpoint, camera lens and the frame acquisition time stamp. This essentially only offers the use of unpaired videos in practice. The evaluation on these two datasets show that our proposed method is capable of achieving highly competitive performance in the user study and by a set of standard benchmark metrics, compared to the state-of-the-art methods. Notably, the proposed method offers significantly better temporal consistency and higher computational efficiency (more than 4x higher than the compared methods) which enables us to perform real-time video enhancement.

In summary, the task of real-time video enhancement with efficient training is achieved by means of several considerations and contributions. In this regard, our main contributions can be summarized as follows:

\begin{enumerate}
    \item We exploit a novel cyclic generative adversarial framework to learn the target distribution from unpaired videos for the task of real-world video enhancement.
    \item We introduce a new recurrent design to both the generators and the discriminator for better spatio-temporal consistency of enhanced videos.
    \item We apply an improved recurrent latent space propagation, for efficient spatio-temporal information diffusion, that highly speeds up the training.
    \item We achieve real-time video enhancement with highly competitive performance both in terms of quantitative and qualitative measures.
    
\end{enumerate}

The rest of the paper is organized as follows.
After setting our work in context to the literature in Sec.~\ref{sec:related_work}, we explain the motivation for our method's design and describe the technical details of the proposed modules in Sec.~\ref{sec:method}. In Sec.~\ref{sec:experiments}, we outline the experimental setup and present the results of our extensive evaluations, comprised of;  quantitative results, a user study, visual examples, temporal consistency curves, and an ablation study. Finally, we conclude our work in Sec.~\ref{sec:conclusion}.

\section{Related Work}
\label{sec:related_work}

The goal of video quality enhancement is to enhance low-quality videos towards those with less artificial noise, more vivid colorization, sharper texture details or higher contrast. In the literature, there are two major families of video quality enhancement methods. The first family aims at focusing on one single subtask of video enhancement. For example, some methods \cite{wang2017novel,dai2017convolutional,yang2018multi} are suggested for video compression artifacts removal, some \cite{liu2010high,varghese2010video,maggioni2012video,godard2018deep,mildenhall2018burst} are designed for video denoising, some \cite{baker2011database,werlberger2011optical,yu2013multi,jiang2018depth,mathieu2015deep,niklaus2017video,jiang2018super,niklaus2018context} are proposed for frame interpolation, and some \cite{su2017deep,aittala2018burst,gast2019deep} are developed for video deblurring. Besides, some \cite{liu2017,duf2018,liu2020end,edvr2019,tao2017,frvsr2018,rlsp2019,chu2018temporally} aim to enhance the resolution of a given video by adding missing high-frequency information. For comprehensive reviews on these topics, we refer readers to \cite{ghoniem2010nonlocal,nasrollahi2014super}. Below we provide the details of four more related works that belong to the first family of video enhancement methods. The first method \cite{xue2019video} uses a neural network with separate trainable motion estimation and a video processing components. Both components are then trained jointly to learn a task-oriented flow, which can be applied to various video enhancement subtasks separately.  
Differently, the second method \cite{chu2018temporally} proposes an GAN-based approach for video super-resolution that results in temporally coherent solutions without sacrificing spatial detail. Similarly, the third method \cite{galteri2019fast} applies the MobileNetV2 architecture \cite{sandler2018mobilenetv2} to the GAN \cite{goodfellow2014generative} model for fast artifact removal from compressed videos. Finally, the method proposed in \cite{xiong2020unsupervised} learns a two-stage GAN-based framework to enhance the real-world low-light images in an unsupervised manner, \cite{lettherebecolor} provides a solution to colorize black-and-white images with full supervision.
More general video domain translation methods are proposed to handle large domain gaps in a broad range of tasks (e.g. \cite{mocycle2019} and \cite{park2019}). 

The most related works, to the method proposed in this work, are the second family of video enhancement methods. These methods aim to strengthen the compound visual qualities of videos, which includes increasing color vividness, boosting contrast, sharpening up textures. The major issue in regard to enhancing videos using the methods of the second family, is the practicality of collecting well-aligned training data. More precisely, these methods require aligned input and target videos, in both the spatial and the temporal domains, which is a challenging requirement to meet in practice. To address this problem, some reinforcement learning based techniques \cite{hu2018exposure,park2018distort,kosugi2019unpaired} create pseudo input-retouched pairs by applying retouching operations sequentially. Other  method \cite{chen2018deep,ignatov2018wespe,ni2020towards} 
exploit GAN models 
for the same task. In this regard, \cite{chen2018deep} suggests to train an image enhancement model by learning the target distribution with unpaired photographs. 
Particularly, the suggested method learns an enhancing map from a set of low-quality photos to a set of high-quality photographs using the GAN technique  \cite{goodfellow2014generative}, which has proven to be good at learning real data distributions. Similarly, \cite{ignatov2018wespe} applies the GAN technique to learn the distribution of separate visual elements (i.e., color and texture) of images. Such separation allows \cite{ignatov2018wespe} to map low-quality images to high-quality with relative ease,  with  more vivid colors and sharpened textures. Although these methods have demonstrated promising success for unsupervised image enhancement, their extension to videos is not trivial besides applying them to frame-by-frame video enhancement. In this work, we show that such straightforward extension produces temporally inconsistent enhancement results. In contrast, due to the well-designed recurrent framework, our proposed method is capable of achieving clearly better temporal consistency as well as much faster inference with highly competitive performance in terms of other benchmark metrics.

\section{Proposed Method}
\label{sec:method}
\begin{figure*}
\begin{center}
\includegraphics[width=0.8\linewidth]{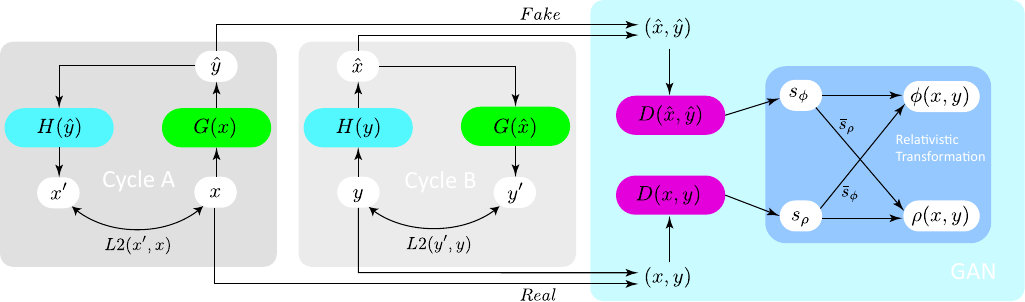}
\end{center}
   \caption{General loss setting. The figure shows two opposite cycles and the single discriminator GAN-loss. A high-quality sequence $\hat{y}=G(x)$ and a low-quality sequence $\hat{x}=H(y)$ are generated from a sample $(x,y)\sim p(x,y)$ drawn from the training data. To favorize a consistent mapping between the domains, generated samples are mapped back and the cyclic loss is evaluated between $x'=H(G(x))$, $y'=G(H(y))$ and the real samples $x,y$. The real $(x,y)$ and fake $(\hat{x},\hat{y})$ samples are fed to our single joint-distribution discriminator $D$ to compute the respective logits $s_\rho$, $s_\phi$. The final relative real $\rho$ and fake $\phi$ scores are computed by applying the relativistic transformation. The GAN-loss is defined as the sigmoid cross-entropy using the sigmoid  $\sigma(\cdot)$.}
\label{fig:objective}
\end{figure*}

One major aim in video enhancement is to strengthen the perceptual quality of videos $X=\{x_1,x_2,\ldots, x_N\}\in \mathbb{R}^{N \times T\times H\times W\times C}$, that are captured on cameras with compromised capabilities, such that they resemble videos $Y=\{y_1,y_2,\ldots,y_M\}\in \mathbb{R}^{M\times T\times H\times W\times C}$ from the high-quality target distribution with the desired properties. 
In addition to the enhancement of stills, video enhancement suffers from new challenges, including; (i) difficulty of collecting spatio-temporally aligned video samples for supervision, (ii) complexity of achieving spatio-temporal consistency within individual videos, (iii) higher computational cost and larger data volumes.

We avoid the problem (i) of collecting aligned pairs by merely learning from unpaired samples in both source and target domain through a
cyclic
GAN-based solution. Our objective function is designed to match the distributions of generated and true samples from the training data alone, without supervision or handcrafted losses. To address the issue of inconsistencies in the temporal and spatial domain (ii), we introduce a fully recurrent framework, consisting of recurrent generator $G$, degrader $H$, and discriminator $D$; all featuring our novel local/global module (LGM). 
We achieve fast training speed and real-time state-of-the-art performance at inference (iii) by employing effective losses and streamlined recurrent architectures, both of which are designed to cope with the heavy computational demand associated with video processing.
In contrast to most generative adversarial frameworks, which require extensive human effort for designing multiple handcrafted losses and the associated expensive tuning of hyperparameters, we address the aforementioned issues from a holistic view of the problem. We shift our design effort from explicit losses to implicit ones, thereby enabling effective learning in a data-driven fashion.
We therefore design a more efficient, data-driven GAN-framework, where only two architectures need to be manually designed, i.e. generator/degrader and discriminator.
The proposed framework features only one recurrent discriminator network (to enforce all desired properties) and leverages joint distribution learning for enhanced coupling between source and target distributions. This leads to  memory- and computation-efficient discrimination in both source domain $\mathcal{X}$ and target domain $\mathcal{Y}$ without the need for separate discriminators. Our discriminator learns the joint distribution $p(x,y)$ directly from unpaired video samples $(x,y)$ in an unsupervised manner.

\subsection{Objective Function}
\label{sec:objective}

Our proposed GAN-framework (Fig.~\ref{fig:objective}) features a novel recurrent generator/degrader and a novel recurrent discriminator to learn the joint distribution $p(x,y)$ using a single network.
With the outputs of discriminator $D$ we calculate real logits $\rho(x,y)$ and fake logits $\phi(\hat{x},\hat{y})=\phi(H(y),G(x))$ from sampled sequences $(x,y)\sim p(x,y)$, which are further evaluated through a standard sigmoid cross-entropy loss.

In addition to the GAN objective, we introduce a cyclic constraint to favorize a bijective mapping between the distributions of source domain $\mathcal{X}$ and target domain $\mathcal{Y}$. The constraint acts mainly as a regularizer to stabilize the training process, which has been successfully applied to unsupervised quality mapping before, e.g. in \cite{chen2018deep}. Our generator $G: \mathcal{X} \rightarrow \mathcal{Y}$ maps videos from source to target domain, while the degrader $H: \mathcal{Y} \rightarrow \mathcal{X}$ constitutes the inverse mapping. In contrast to previous methods, we are left with only one single hyperparameter $\alpha$ to balance the GAN-loss with the cyclic constraint and thus significantly reduce the burden on tuning hyperparameters. For enhanced coupling two symmetric cycles are employed; one starting from the source sequence $H(G(x))=x'$, and the other starting from the target sequence $G(H(y))=y'$. 
Note, contrary to recently proposed GAN training strategies, our setting requires no normalization or advanced regularization techniques \cite{miyato2018spectral,gulrajani2017wgangp} for stable training, even with a batch size of only 1 sample. 
Generator $G$, degrader $H$ and discriminator $D$ are trained in alternating fashion:

\begin{equation}
\label{eq:gan_objective}
    \begin{aligned}
    \min_{G,H}\text{ } & \mathcal{L}^G_{GAN}\left(G,H,D\right) + \alpha\mathcal{L}_{cyc}\left(G,H\right), \\ 
    \min_{D}\text{ } & \mathcal{L}^D_{GAN}\left(G,H,D\right),
    \end{aligned}
\end{equation}
where the above three loss terms are given by 
\begin{equation}
\label{eq:lossWithPhiRho1}
\begin{aligned}
\mathcal{L}^G_{GAN}\left(G,H,D\right) = & -\mathbb{E}_{(x,y)\sim p(x,y)}\left[\log(\sigma\left(\phi(x,y)\right)\right] \\
& - \mathbb{E}_{(x,y)\sim p(x,y)}\left[\log(1 - \sigma\left(\rho(x,y)\right)\right],
\end{aligned}
\end{equation}
\begin{equation}
\label{eq:lossWithPhiRho2}
\begin{aligned}
\mathcal{L}^D_{GAN}\left(G,H,D\right) = & -\mathbb{E}_{(x,y)\sim p(x,y)}\left[\log(\sigma\left(\rho(x,y)\right)\right] \\
& - \mathbb{E}_{(x,y)\sim p(x,y)}\left[\log(1 - \sigma\left(\phi(x,y)\right)\right],
\end{aligned}
\end{equation}
\begin{equation}
\label{eq:cyclic}
\begin{aligned}
\mathcal{L}_{cyc}\left(G,H\right) = \mathbb{E}_{(x,y)\sim p(x,y)}\left[(x' - x)^2 + (y' - y)^2\right].
 \end{aligned}
\end{equation}

We use a relativistic average GAN objective (RaGAN) \cite{jolicoeur2018relativistic} to train our networks,  since it has been shown to be  effective for enhancement tasks~\cite{wang2018esrgan}.
We however adapt  RaGAN to train using a single sample per batch,  by avoiding  the original averaging performed over the batch dimension. In memory constrained GPUs, our single sample-based training enables larger crop sizes, which was observed to be important for video enhancement with global properties. 
The discriminator $D$ accepts real $(x,y)$ and fake $(\hat{x},\hat{y})$ samples and outputs the respective logits $s_\rho,s_\phi\in\mathbb{R}^{T\times H\times W\times 1}$ for each pixel-location. 
Then, we average over the map of logits $s_\rho = D(x,y), s_\phi = D(\hat{x},\hat{y}) \in \mathbb{R}^{T\times H\times W\times 1}$ in the spatial domain. 
Analogous to \cite{jolicoeur2018relativistic}, this adaption leverages the fact that all patches in a frame should be classified exclusively as real or fake.
The averages $\overline{s}_\rho, \overline{s}_\phi$, obtained as follows, can then be used for the relativistic transformation. 

\begin{equation}
    \overline{s}_\rho=\frac{1}{HW}\sum_{h}^{H}\sum_{w}^{W} s_\rho(t,h,w),
\end{equation}
\begin{equation}
    \overline{s}_\phi=\frac{1}{HW}\sum_{h}^{H}\sum_{w}^{W} s_\phi(t,h,w).
\end{equation}
After the relativistic transformation, the modified logits for real $\rho(x,y)$ and fake $\phi(x,y)$ are given by,

\begin{equation}
\label{eq:rho}
\begin{aligned}
\rho(x,y) = & D(x,y)-\overline{s}_\phi,
 \end{aligned}
\end{equation}
\begin{equation}
\label{eq:phi}
\begin{aligned}
\phi(x,y) = & D(\hat{x},\hat{y}) - \overline{s}_\rho.
 \end{aligned}
\end{equation}
These modified logits are  then evaluated with the standard sigmoid cross-entropy loss, using the loss functions ~\eqref{eq:lossWithPhiRho1} and ~\eqref{eq:lossWithPhiRho2}. Please, refer to Fig.~\ref{fig:objective} for our loss setting.

\begin{figure}[t]
\begin{center}
\includegraphics[width=.95\linewidth]{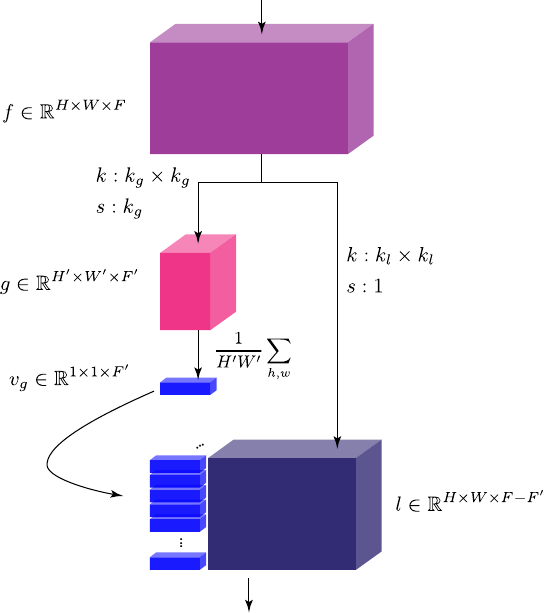}
\end{center}
   \caption{LGM module. A standard convolution branch (right) processes local information, while a second branch (left) extracts global information from the input feature map $f$. The features $g$, computed from $f$ with kernel size $k_g$ and stride $k_g$, are average pooled to form the global feature vector $v_g$. This vector is replicated spatially and concatenated with the local features $l$. Adding the LGM module to our model, increases the runtime only marginally by 2.5ms, but improves the performance significantly.}
\label{fig:lgm}
\end{figure}

\begin{figure*}[t]
\begin{center}
\includegraphics[width=0.95\textwidth]{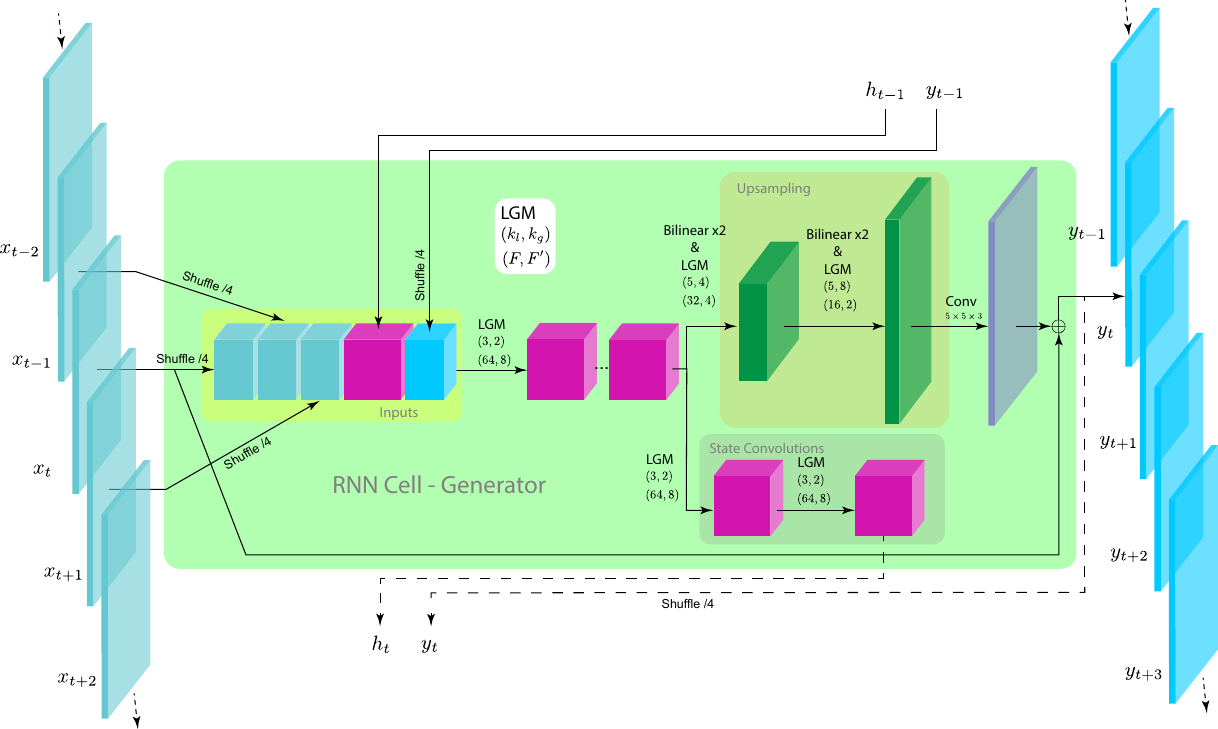}
\end{center}
   \caption{Spatio-temporal generator. The figure shows the recurrent cell at time step $t$. From a batch of input frames $x_{t-1}, x_t, x_{t+1}$, hidden state $h_{t-1}$ and feedback $y_{t-1}$ a single enhanced frame $y_t$ is produced. The employment of our novel LGM block enables processing of global spatial information, $k_l,k_g$ denote the kernel sizes for local and global feature maps, $F$ denotes the total number of filters in a LGM block, $F'$ is the number of filters in the global feature maps. These parameters are set according to their respective spatial input sizes, see Sec.~\ref{sec:network_design} and Fig.~\ref{fig:lgm}.
   The residual connection from input $x_t$ to output $y_t$ enables learning only the residuals with respect to the input center frame $x_t$, $\oplus$ denotes element-wise tensor addition.}
\label{fig:generator}
\end{figure*}

\begin{figure*}[ht!]
\begin{center}
\includegraphics[width=0.95\textwidth]{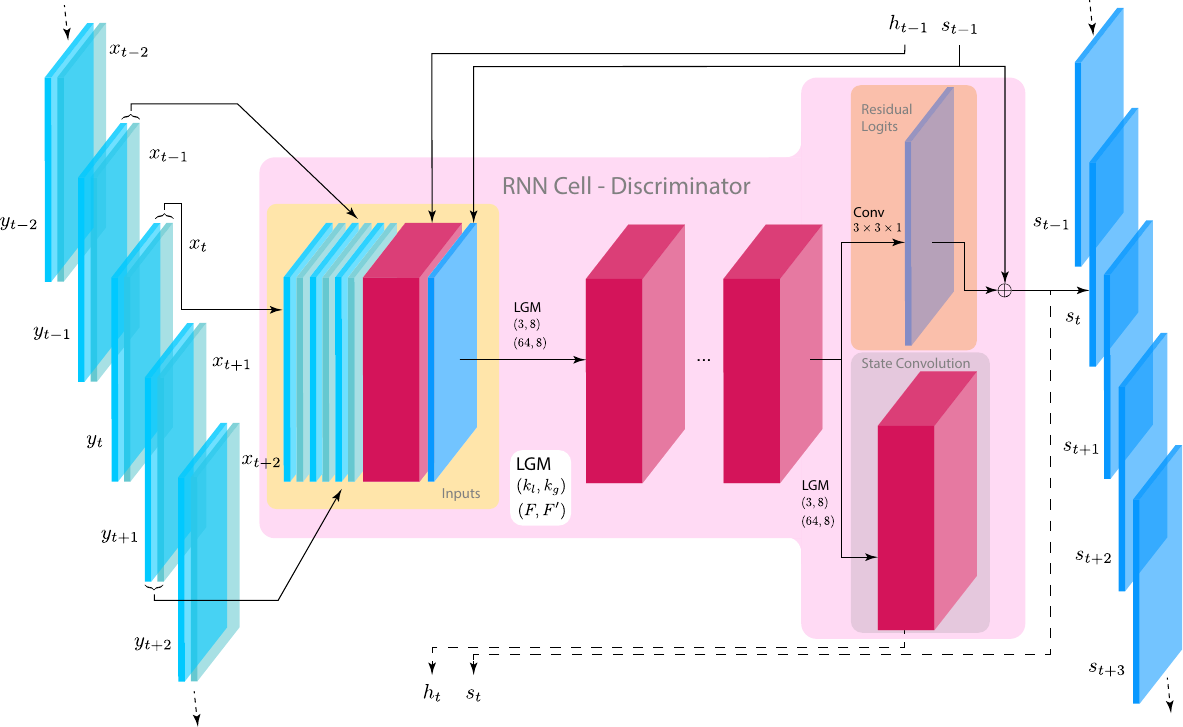}
\end{center}
   \caption{Spatio-temporal discriminator. 
   The figure shows the recurrent cell at time step $t$. As the discriminator is designed to learn the joint distribution of source $\mathcal{X}$ and target $\mathcal{Y}$ domain, sequences $x,y$ are concatenated along the channel dimension and fed to the discriminator.
   A map of logits $s_t$ is produced from a batch of concatenated input frames,
   hidden state $h_{t-1}$ and feedback $s_{t-1}$. The employment of our novel LGM block enables processing of global spatial information, $k_l,k_g$ denote the kernel sizes for local and global feature maps, $F$ denotes the total number of filters in a LGM block, $F'$ is the number of filters in the global feature maps, see Sec.~\ref{sec:network_design} and Fig.~\ref{fig:lgm}.
   Note, we compute residual logits in relation to the previous map of logits $s_{t-1}$, $\oplus$ denotes element-wise tensor addition.}
\label{fig:discriminator}
\end{figure*}

\subsection{Architecture Design}
\label{sec:network_design}

In order to achieve fast runtimes, temporally consistent video, and strong training signals, we modify RLSP~\cite{rlsp2019}, a powerful recurrent architecture for supervised real-time video super-resolution (VSR). Our modifications allow us to perform video quality mapping (VQM) with fully convolutional recurrent cells. Since every component in our setting benefits from recurrence, all networks ($G$,$H$ and $D$) are derived from RLSP. Recurrence allows for consistency in the temporal domain, as the processing at the current time step is constrained with the previous time step’s hidden state and it enables efficient temporal information accumulation in all architectures.
In addition to a local view for improvement of properties like sharpness and contrast, VQM benefits from a global view of the input in order to perform spatially coherent enhancement, e.g. overall illumination and tone. For this purpose, we introduce a novel local and global module (LGM) within both of the generators and the discriminator. 
Combined application of recurrence and LGM establishes a fully connected view from all the past spatial and temporal information in a sequence to individual pixels in the output, improving spatio-temporal consistency. Futhermore, such a design suits online inference with our enhancer $G$, as it facilitates the maximum usage of available information at each time step.

\noindent\textbf{Local/Global Module (LGM):} Our introduced LGM block for efficient interleaved local and global information processing consists of two convolutional blocks. In addition to a standard local convolution block with a $k_l\times k_l$ kernel, a second block with a larger $k_g\times k_g$ kernel with stride $k_g$ extracts features from equidistant patches. Features $g\in \mathbb{R}^{H'\times W'\times F'}$ from this second block are average pooled in the spatial domain to obtain a global feature vector $v_g \in \mathbb{R}^{1\times 1\times F'}$. 
\begin{equation}
v_g=\frac{1}{H'W'}\sum_{h,w}^{H',W'} g(h,w,f)
\end{equation}
Then, the pooled features are spatially replicated to match the previous input size and concatenated with the local feature maps $l\in \mathbb{R}^{H\times W\times F-F'}$ (global concatenation), acting as global priors for the downstream convolutions, see Fig.~\ref{fig:lgm}.

Repeated LGM blocks enable interleaved processing from local to global information. Note, after only one processing step with the LGM module, a fully connected global view is established, which  dynamically adapts to the spatial size of the input. Subsequent processing steps further refine the global and local relations in multiple stages and establish a rich set of features containing information from the full global view all the way down to pixel-level.

\subsubsection{Generator}
\label{sec:generator}

We introduce the LGM module as a replacement convolution block in RLSP to efficiently process local and global information with the help of recurrent connections.
The proposed recurrent cell features a hidden state and feedback, which enables the architecture to produce temporally consistent video enhancement.
Furthermore, a shuffle operation ($/4$) at the input increases the receptive field and enables efficient computation by reducing spatial size. At the output stage, the features are upscaled to the original resolution through combined bilinear resizing and convolution, yielding the enhanced frames $\hat{y}=G(x)$. In each upsampling stage, the number of filters in the LGM modules ($F,F'$) is divided by 2. The kernel sizes $k_g$ in the LGM modules are set according to the current spatial resolution in relation to a full-resolution kernel size of $k_g=8$. In the first blocks after the shuffling operation ($/4$) we set $k_g=8/4=2$, in the subsequent two upscaling blocks we set $k_g=8/2=4$ and $k_g=8$ respectively. 
Two extra LGM blocks are placed in parallel to the upsampling block for further refinement in the hidden state. The number of layers are set to 8. The resulting architecture allows joint processing of spatio-temporal as well as global and local information in a single pipeline. We initialize the values for feedback and hidden state $f^0,h^0$ with zeroes. We use the same network for the degrader $H$ to impose the cyclic constraint of Eq.~\ref{eq:cyclic}.

\subsubsection{Discriminator}

Our recurrent Discriminator architecture also incorporates the LGM module, which is again based on RLSP. Including recurrence to the discriminator adheres to introducing a suitable bias for spatio-temporal data. With the  similar considerations as that of the generator, recurrence encourages to discriminate temporal dynamics. This efficiently improves the capacity of the discriminator, leading to stronger training signals for the generator.  Here again, the used LGM blocks exploit local and global information, which plays a vital role in VQM.

The architecture of the proposed network is a fully convolutional recurrent one, which operates in full resolution, see Fig.~\ref{fig:discriminator}. The cell is implemented with 8 layers. We predict pixel-dense logits $s \in \mathbb{R}^{T\times H\times W\times 1}$ per sequence which are fed to the GAN objective after applying the relativistic transformation, see Eq.~\ref{eq:rho} and \ref{eq:phi}. Due to the repeated LGM blocks and recurrence, every local pixel classification in $s$ also includes the full past spatio-temporal information. Since all operations are conducted in full resolution, we set the global kernel size to $k_g=8$ in order to be consistent with the choice of $k_g$ in the generator. The discriminator also features a hidden state and feedback. The logits $s$ are fed back to the input by concatenation with the inputs and the hidden state, serving as priors at the next time step. The network predicts residual logits $s'_t\in \mathbb{R}^{H\times W\times 1}$, which are added to the previous output $s_{t-1}$ to form the logits $s_t$ at the current time step $t$. The hidden state propagates 
information along time and thereby enables the discrimination of temporal dynamics. Along with discrimination in the spatial domain, our recurrent discriminator can also enforce realistic temporal dynamics over long sequences. Note that the latter  is not the case for non-recurrent discriminator networks.

\begin{figure}[t]
\begin{center}
\includegraphics[width=0.95\linewidth]{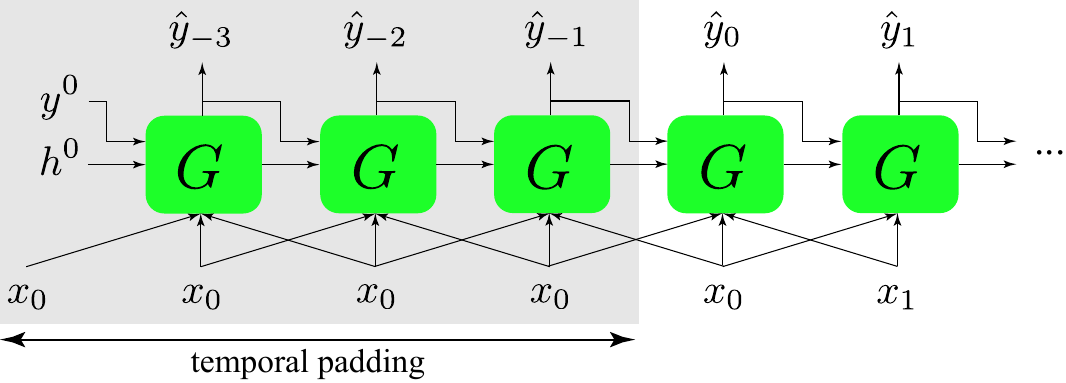}
\end{center}
   \caption{Temporal padding, shown for the generator. The first frame is replicated and prepended before processing any sequence. This enables initialization of hidden state $h_{t-1}$ and feedback $\hat{y}_{t-1}$ without the pressure of generating meaningful output right away. The padding is removed after processing, i.e. $\hat{y}_{-3}, \hat{y}_{-2}$ and $\hat{y}_{-1}$ are removed.}
\label{fig:padding}
\end{figure}

\section{Experiments and Discussion}
\label{sec:experiments}

\subsection{Experimental Setup}
To get manageable training samples, we crop the videos online in the spatio-temporal domain to short sequences $x,y\in\mathbb{R}^{5\times256\times256\times3}$. Since our proposed recurrent adversarial video enhancement (RAVE) framework\footnote{Codes, links to data and other material will be available at: \url{https://github.com/dariofuoli/RAVE}} is recurrent, hidden state and feedback need to be initialized at the beginning of a sequence, both $h^0$ and $y^0$ are set to zero. 
In addition, the efficiency gain in our proposed recurrent networks stems from access to past information which is encoded in the current hidden state. This information is not available at the beginning of a sequence, as it needs to first accumulate for a couple of frames. Therefore, the enhancement quality gradually increases during the first few frames, which is a discriminating factor for the discriminator by itself and also affects the cyclic losses. 
In order to prevent initialization-related issues during training, all sequences are padded by repeating the first frame for 3 time steps, see Fig.~\ref{fig:padding}.
At every step during training, i.e. after a sequence is processed, the excess frames are removed. Therefore, the padded frames are never exposed to any losses directly, leaving room for the generator G and degrader H to warm-start the hidden state.
The proposed temporal padding strategy alleviates the need for constant enhancement quality from the first frame on and enables the network to make use of past information without exposing the initialization process to the discriminator.
We observed improved stability during training and better training results with this modification.
We use Adam optimizer \cite{adam} with standard settings from \textit{PyTorch} and set the learning rate to $10^{-6}$. Due to hardware constraints, all models are trained with a batch size of 1. The cyclic loss weight is set to $\alpha=1$ for all experiments,  generator/degrader and discriminator are trained in alternating steps.

\subsection{Datasets}
\label{sec:datasets}
We use two FullHD ($1920\times1080$) video datasets to evaluate and compare our proposed RAVE method against competitors. Vid3oC \cite{kim2019vid3oc} is a small dataset consisting of 50 roughly-aligned videos for training and 16 videos for validation recorded with 3 different cameras (a ZED stereo camera, a Canon 5D Mark IV, and a Huawei P20) on a calibrated camera rig. This dataset can be publicly downloaded and has been used in challenges for video super-resolution \cite{AIM2019VXSRchallenge,fuoli2020aim_VXSR} and video quality mapping \cite{fuoli2020ntire}. Following the setup of \cite{fuoli2020ntire}, we use the videos taken by the ZED (low-quality i.e., source) and Canon (high-quality i.e., target) cameras for the video enhancement evaluation. The data consists of a training set with 50 source videos and 50 target videos, as well as a validation set\footnote{Since the full test set is not released to the public, we merely use the validation set for evaluation.} with 16 source videos and 16 target videos. Alignment is not required by our framework, nevertheless we use this dataset as it was used as benchmark in \cite{fuoli2020ntire} and it allows us to compute approximate LPIPS scores, which rely on aligned frames. We completely ignore the fact, that the videos are roughly aligned and sample $x,y$ independently during training. However, there are several issues attached to this dataset. Good generalization to the validation set from learning on only 50 training sequences is highly doubtful as the sample size is simply too small for video. On top of that, the target domain contains serious motion blur, which is not the desired property in video enhancement and causes the generator to produce blurred frames.

To address these shortcomings we employ a large dataset with videos from a smartphone camera (Huawei P30 Pro) and a high-quality DSLR (Panasonic GH5s), which we call Huawei. Videos are captured by the P30 Pro in 29.95fps, while those taken by the Panasonic GH5s are of 25fps. The resolution of the captured videos is also $1920 \times 1080$. The videos for both devices have been shot at the same locations, but without proper alignment in any dimension. The viewpoints, lenses and recording times are very different. The data is split into 1270 source videos and 1270 target videos for training, and 16 pairs of validation source/target videos. Again, we sample $x,y$ independently.

\subsection{Evaluation}
\subsubsection{Quantitative Metrics}
The design of reliable quantitative metrics for unsupervised generation and enhancement, which correlate well with human perception, is difficult and is still ongoing research. Missing pixel-level aligned targets in these settings prohibit the use of standard distance metrics like PSNR and SSIM and even more recent feature-level based metrics like LPIPS \cite{zhang2018perceptual} are no option when source and target are unaligned. Nevertheless, we provide approximate LPIPS scores on the Vid3oC dataset, as the frames are roughly aligned. Due to the severe misalignment, no meaningful LPIPS scores can be provided on the Huawei dataset. 

A popular metric to measure the distance between two distributions on image level is the Fréchet Inception Distance (FID) \cite{Heusel2017fid,obukhov2020torchfidelity}. FID represents the Fréchet distance between generated samples and samples from the target distribution in feature space of Inception v3 \cite{Szegedy2016inceptionv3}. We calculate FID scores on all frames in the validation set to compare frame-level mapping quality of our method. Although this metric is suitable for our purpose, it is restricted to measure frame-level quality and neglects the temporal part of our video distributions entirely. We therefore also compute the video based FID (i.e., FVD) scores, proposed in \cite{unterthiner2018fvd}. Contrary to FID, FVD computes the Fréchet distance from video based features, which ultimately enables to measure the distance between two video distributions.
\subsubsection{Qualitative Metrics}
Furthermore, a user study is conducted to support the metrics. For that matter we asked the users to mark their preference of the mapping quality towards the target distribution between two methods. In each comparison, the enhanced frames from the two competing methods were shown together with a frame from the target sequence. The users were asked to rate which method enhances the image features closer to the target. Together with the frames we also provided a zoomed-in crop for low-level inspection. One frame per validation video sequence is randomly extracted for comparison to be rated by 20 users, resulting in $16\times 20 = 320$ ratings. To rank all methods among each other, we calculate the total number of wins against the competing methods, normalized by the total number of comparisons ($2\times 320 = 640$). This ratio is given as percentage in Tab.~\ref{tab:results}.
We also provide visual examples for results on the validation sets of the Vid3oC and Huawei datasets in Figs.~\ref{fig:visual_results_vid3oc},\ref{fig:visual_results_huawei}.

\subsubsection{Speed Metric}
To compare the methods in terms of inference speed, we compute runtimes on an NVIDIA V100. The runtimes are measured per frame with equal settings on a FullHD ($1080\times1920$) sequence from the validation set. Since we use temporal padding during training and inference for RAVE, the processing time for padded frames are also considered and therefore slightly increase the runtimes for RAVE. The best out of 10 runs is reported for each method.

\subsubsection{Temporal Consistency Metric}
\label{sec:temporal_consistency}
Current metrics for perceptual quality mostly neglect the temporal domain entirely. Consistent enhancement along time however is crucial for high-quality videos, as the human vision is sensitive to inconsistencies like flickering artifacts. Due to the lack of reliable temporal consistency metrics for unsupervised settings in the field, we propose a novel way of quantifying our model as well as the competing methods with respect to temporal consistency. 
Since there is no pixel-level ground truth in unsupervised video enhancement, it is difficult to obtain such a metric with reference to the target domain. However, source $x$ and enhanced video $\hat{y}=G(x)$ are aligned on pixel level. With the assumption that the source video is temporally consistent, since we do not expect temporal inconsistencies in recordings, we can use it as reference. 

Pixel-flow for source $F_{S,t}(h,w)=x_{t+1}(h,w) - x_t(h,w)$ and method $F_{M,t}(h,w)=\hat{y}_{t+1}(h,w) - \hat{y}_t(h,w)$ are calculated per sequence. In order to compare the method's flows against the reference $x$, we calculate the statistics from the flow difference $d_t = F_{M,t}(h,w) - F_{S,t}(h,w)$ per time step, which we call relative pixel-flow (RPF), and plot it over time.
\begin{align}
\mu_{d_t} &= \frac{1}{HW}\sum_{h}^{H} \sum_{w}^{W} F_{M,t}(h,w) - F_{S,t}(h,w) \\
\sigma_{d_t} &= \sqrt{\frac{1}{HW}\sum_{h}^{H} \sum_{w}^{W} \left[F_{M,t}(h,w) - F_{S,t}(h,w) - \mu_{d_t} \right]^2}
\end{align}
The mean flow $\mu_{d_t}$ and standard deviation $\sigma_{d_t}$ for the first 8 sequences from both validation sets are plotted in Figs.~\ref{fig:rpf_vid3oc},\ref{fig:rpf_huawei} for all methods. 
A temporally consistent generated video $\hat{y}$ should have similar pixel-flow as the source video. The similarity is indicated by a small mean $\mu_{d_t}$ and standard deviation $\sigma_{d_t}$, together with smooth evolution of $\mu_{d_t}$ and $\sigma_{d_t}$ over time, e.g. no spikes and no valleys.

\begin{table*}
  \caption{
  Comparison with state-of-the-art. We conducted a user study and compute FID, FVD, LPIPS and runtime to evaluate the mapping quality and the performance on the Vid3oC \cite{kim2019vid3oc} and Huawei datasets. \textit{Wins} denotes the normalized relative preference to both other methods in \%.}
  \label{tab:results}
  \centering
  \begin{tabular}{llcccrrrrrr}
    \toprule
    
    && \multicolumn{4}{c}{User Study} & \multicolumn{4}{c}{Metrics}\\[4pt]
    & Method & WESPE & DPE & RAVE & Wins & $\downarrow$FID & $\downarrow$FVD  & $\downarrow$LPIPS & 
    $\downarrow$Runtime [ms] \\
    \midrule
     \multirow{3}{*}{\rotatebox{90}{Vid3oC}} 
     & WESPE  & - & 173/147 & 161/159 & \textbf{52\%} & 52.66 & 1068.88 & 0.584 &  284.6 \\
     & DPE & 147/173 & - & 154/166 & 47\% & 56.10 & 1048.66 & 0.631 &  117.9 \\
     & RAVE (ours) & 159/161 & 166/154 & - & 51\% & \textbf{51.85}  & \textbf{968.04} & \textbf{0.578} &   \textbf{27.8} \\
     \midrule
     \multirow{3}{*}{\rotatebox{90}{Huawei}}
     & WESPE & - & 134/186 & 100/220 & 37\% & 115.84 & 2134.07 & n/a &  284.6 \\
     & DPE & 186/134 & - & 142/178 & 51\% & 115.46 & 2125.66 & n/a &  117.9 \\
     & RAVE (ours) & 220/100 & 178/142 & - & \textbf{62\%} & \textbf{113.54}  & \textbf{2099.39} & n/a &  \textbf{27.8} \\
     
    \bottomrule
  \end{tabular}
\end{table*}

\begin{figure*}[h!]
\begin{center}
\includegraphics[width=0.78\linewidth]{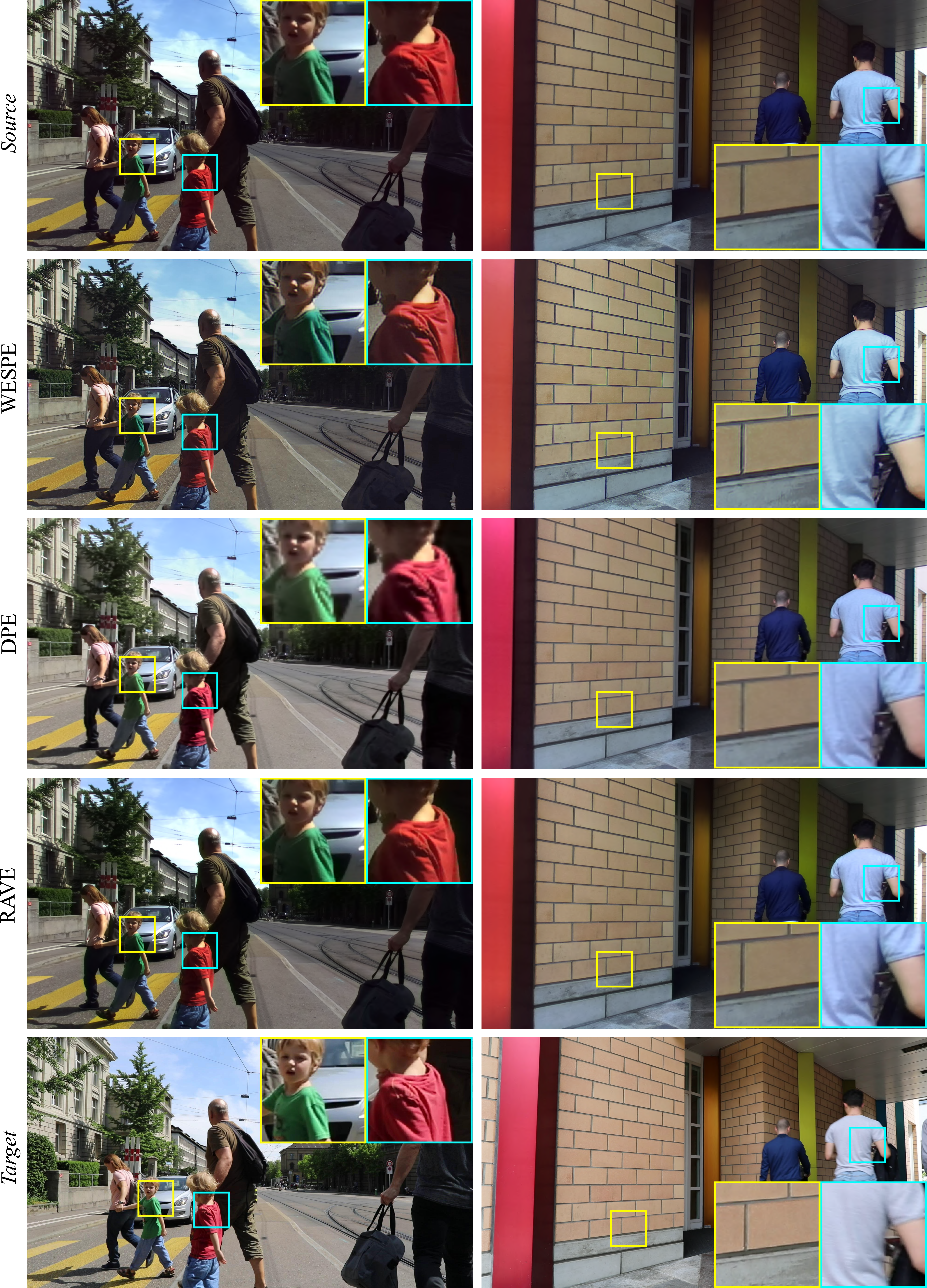}
\end{center}
   \caption{Visual results for Vid3oC.}
\label{fig:visual_results_vid3oc}
\end{figure*}

\begin{figure*}[h!]
\begin{center}
\includegraphics[width=0.78\linewidth]{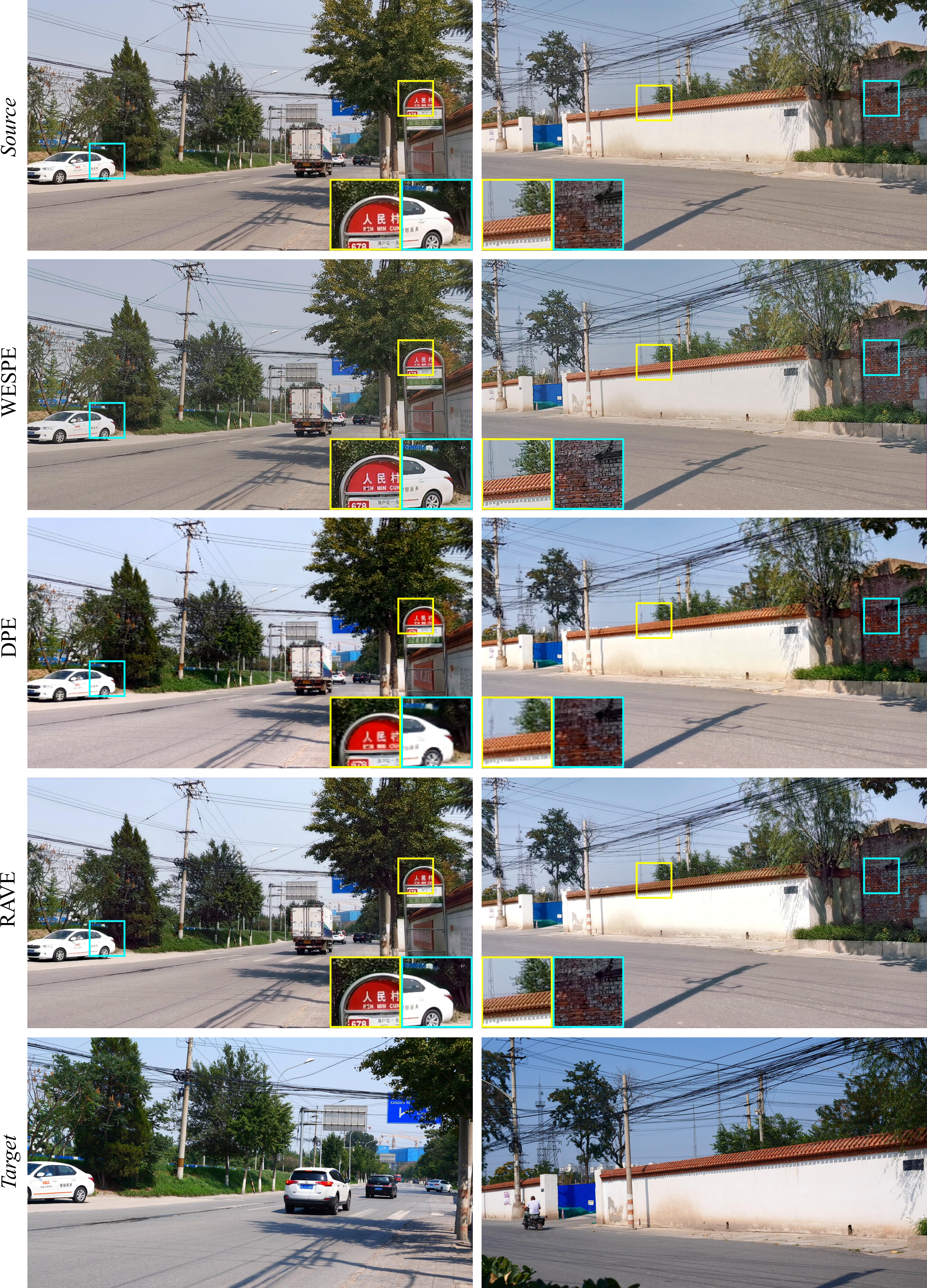}
\end{center}
   \caption{Visual results for Huawei. We do not highlight details for the target, due to different view, scale and time.} 
\label{fig:visual_results_huawei}
\end{figure*}

\subsection{Comparison with State-of-the-Art Methods}

\begin{table*}[ht]
  \caption{Ablation study. We show the effectiveness of our introduced modules on the Vid3oC dataset \cite{kim2019vid3oc} in terms of FID, FVD and LPIPS scores.}
  \label{tab:ablation}
  \centering
  \begin{tabular}{ccccrrrrr}
    \toprule
    Config. &Recurrence    & RaGAN* & LGM & FID$\downarrow$ & $\downarrow$FVD & $\downarrow$LPIPS & $\downarrow$Runtime [ms] \\
    \midrule
    1&&&& 60.91 & 1048.91 & 0.582 &\textbf{22.3}\\
    2& x &&& 60.36 & 1092.26 & 0.583 &25.3\\
    3 && x &  &  60.63  & 1067.17 & 0.582 & 22.3 \\
    4 &x & x & & 59.10  & 1049.48 & 0.583 & 25.3 \\
    5 &&  x & x & 53.37 & 983.13 & \textbf{0.578} & 24.3\\     
    6 &x & x & x & \textbf{51.85} & \textbf{968.04} & \textbf{0.578} & 27.8 \\
    \bottomrule
  \end{tabular}
\end{table*}

To the best of our knowledge, existing published works rarely apply unsupervised quality enhancement to videos.
So far, only methods for image level quality mapping exist. We therefore compare our proposed model RAVE to the most prominent state-of-the-art single image enhancers in the literature for quality mapping, Weakly-Supervised Photo Enhancement (WESPE)~\cite{ignatov2018wespe} and Deep Photo Enhancement (DPE)~\cite{chen2018deep}. WESPE divides the enhancement problem into two subtasks and treats the texture and color enhancement separately and features a cyclic constraint which is enforced through a VGG loss \cite{Ledig_2017_CVPR,Simonyan2015vgg}. DPE uses a cyclic GAN setting with identity loss, cyclic constraint and two GAN losses, one for each domain. 
The major difference to our proposed method is the counter productive identity loss, the lack of taking into account the temporal properties for smooth videos and neglecting the importance of fast runtimes to enable real-time applications.
Besides, the network is not fully convolutional, which inherently limits its practical application due to the fixed input size. For both competing methods, we employ the implementations from the original authors.
We evaluate quantitative metrics and compare the perceived visual quality by providing examples and a user study.

We additionally compare our model against a method for video translation~\cite{Recycle-GAN}, that shares similarities to the VQM task (see supplementary material). The method is unable to produce comparative quantitative (FID of 56.68 on Vid3oC) and qualitative results, which we attribute to the differences between both tasks. The results highlight the need for a specialized solution for VQM, which we provide with our proposed framework.

\subsubsection{Quantitative Results}
The results on both datasets are listed in Tab.~\ref{tab:results}. We outperform the current state-of-the-art methods DPE and WESPE in all quantitative metrics on both image level (FID, LPIPS) and video level (FVD). 
We gain up to 4.25 (Vid3oC) and 2.30 (Huawei) in FID, achieve better FVD scores up to 100.84 (Vid3oC) and 34.68 (Huawei) and up to 0.053 lower LPIPS values over both competing methods WESPE and DPE.
Despite the (implicit) additional temporal consistency constraint imposed by our recurrent discriminator, our method produces higher quantitative results on image level, which is quite remarkable as both image level enhancers are not bound to this constraint and can therefore solve a simpler objective.
Our RAVE clearly outperforms the other methods in terms of FVD scores due to the recurrent discriminator and also achieve the fastest runtimes. Our proposed RAVE method is the only method that achieves real-time performance by producing over 35fps of high-quality FullHD ($1080\times1920$) video. RAVE is over $\times 10$ faster than WESPE and over $\times 4$ faster than DPE. DPE only produces video in a reduced resolution ($512\times512$) and therefore has to process about $\times 7.9$ less pixels than WESPE and RAVE. Hence, the speed difference to RAVE would be larger, if RAVE would process the same resolution. Please note, high-quality real-time performance is generally hard to achieve in deep learning, mainly because designing the optimal network structure is crucial when its model complexity should be minimized.

\subsubsection{Qualitative Results}

In addition to the quantitative metrics we provide visual examples to inspect the perceived visual quality. We show a frame-level comparison between the unaltered source frame, WESPE, DPE, RAVE and a corresponding target frame from the validation set of Vid3oC in Fig.~\ref{fig:visual_results_vid3oc}. Note, the temporal consistency is covered later in Sec.~\ref{sec:temporal_consistency_results}. As previously stated in Sec.~\ref{sec:datasets}, the dataset size is limited for the task. DPE can match the colors a bit closer to the target, which however comes at the expense of a significantly lower resolution due to the inevitable resizing operations to facilitate the fixed input size of the network. The results are therefore seriously blurred out, as can be seen by the close-ups. Effectively, DPE is trained on complexity-reduced data due to the resampling into the lower resolution. The resulting lower Nyquist rate, with its associated lack of higher frequencies,  removes the task of matching the sharpness of generated and target frames and reduces the problem to color matching only. Additionally, the model is learned on the full spatial content, which improves generalization when compared to learning on smaller crops only.
WESPE produces high frequencies which perceptually sharpens the output. However, a closer look reveals low-level artifacts, e.g. discontinuous lines and ringing artifacts on the edges. Additionally, there is a pattern of bright, green pixels distributed over the whole frame, which appears in all outputs. RAVE produces balanced frames without those artifacts and achieves a better trade-off between color and sharpness.

\begin{figure}[t]
\begin{center}
\includegraphics[width=\linewidth]{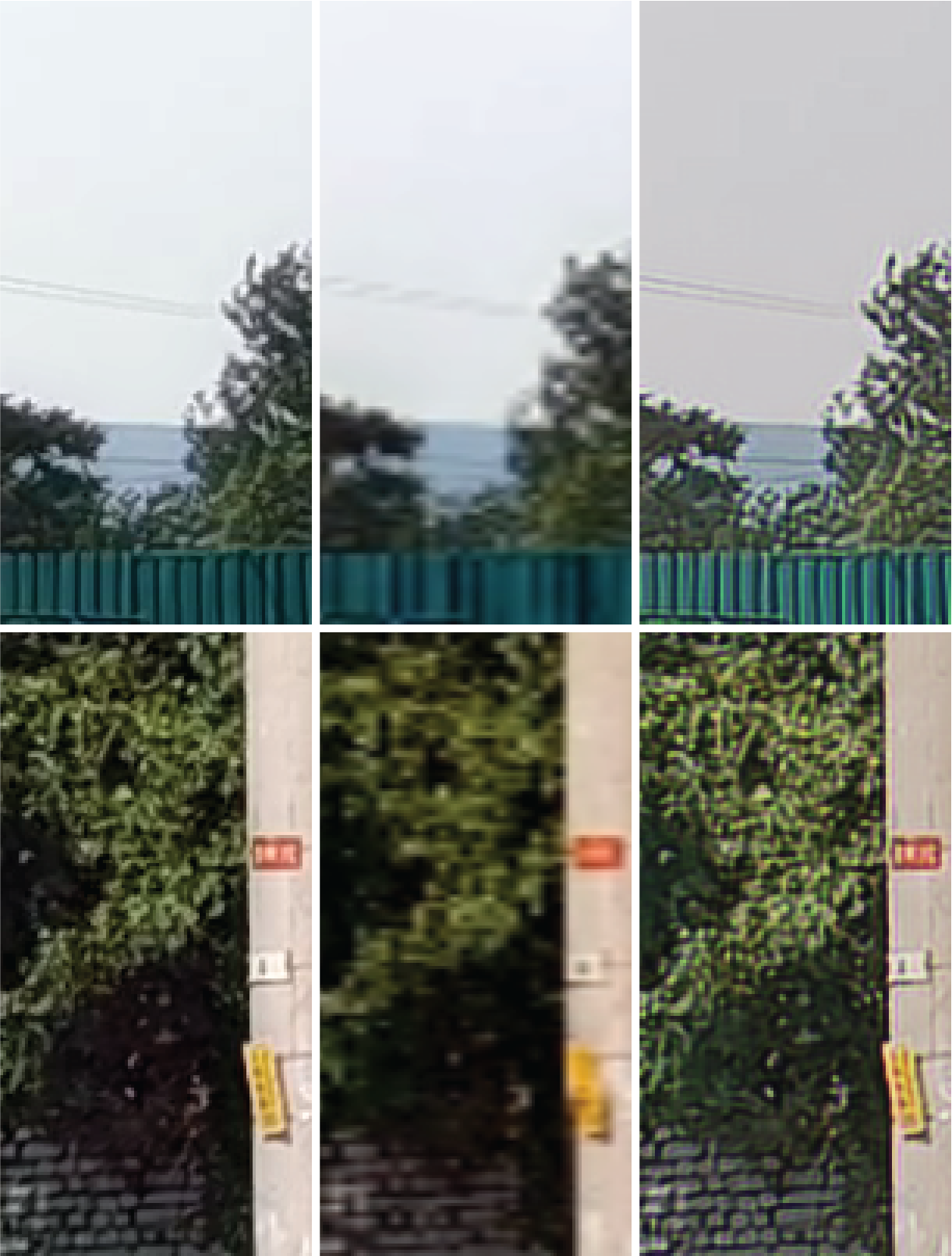}
\end{center}
   \caption{Close-up visual results of RAVE (left), DPE (center) and WESPE (right) for Huawei. Note the blurry content of DPE and the heavy low-level artifacts (green) of WESPE. Additionally, WESPE is missing the desired color properties from the target domain. RAVE produces clean frames with properties similar to the target domain.}
\label{fig:close_ups_huawei}
\end{figure}

The results on the larger Huawei dataset in Fig.~\ref{fig:visual_results_huawei} and additional close-ups in Fig.~\ref{fig:close_ups_huawei} are more representative due to the aforementioned discussion in Sec.~\ref{sec:datasets}. RAVE achieves significantly better color matching than on Vid3oC and is on par or better than DPE in this regard, despite the more complex requirements for sharpness and temporal consistency. Again, WESPE produces higher frequencies, but the exact same low-level artifacts can be observed as on Vid3oC. Also, WESPE fails completely to learn the target color distribution and generates colors which are close to the source, but seem even further away from the target. Additionally, it does not pick up the contrast from the target distribution, which RAVE can handle impressively. RAVE attains good color matching, higher sharpness compared to DPE and can process the data in real-time.

The user study reflects the aforementioned issues associated with each method on the Vid3oC dataset, since no method can match the target distribution satisfactorily. Hence, the preference scores among the users are indecisive. RAVE outperforms DPE and is on par with WESPE, separated by only one vote, while WESPE achieves better scores against DPE.
RAVE is the clear winner on the Huawei dataset. RAVE's video enhancement is significantly preferred over both DPE and WESPE, as it wins 62\% of all direct comparisons. The user study's results align well with our visual quality discussion above, that yielded significant issues for WESPE. WESPE is favored only in 37\% of all direct comparisons. RAVE is preferred over WESPE with a ratio of $220/100$, and outperforms DPE with a ratio of $186/134$.

\subsubsection{Temporal Consistency Results}
\label{sec:temporal_consistency_results}

\begin{figure*}[ht!]
\begin{center}
\textbf{RPF Curves - Vid3oC}\par\medskip
\includegraphics[width=0.85\linewidth]{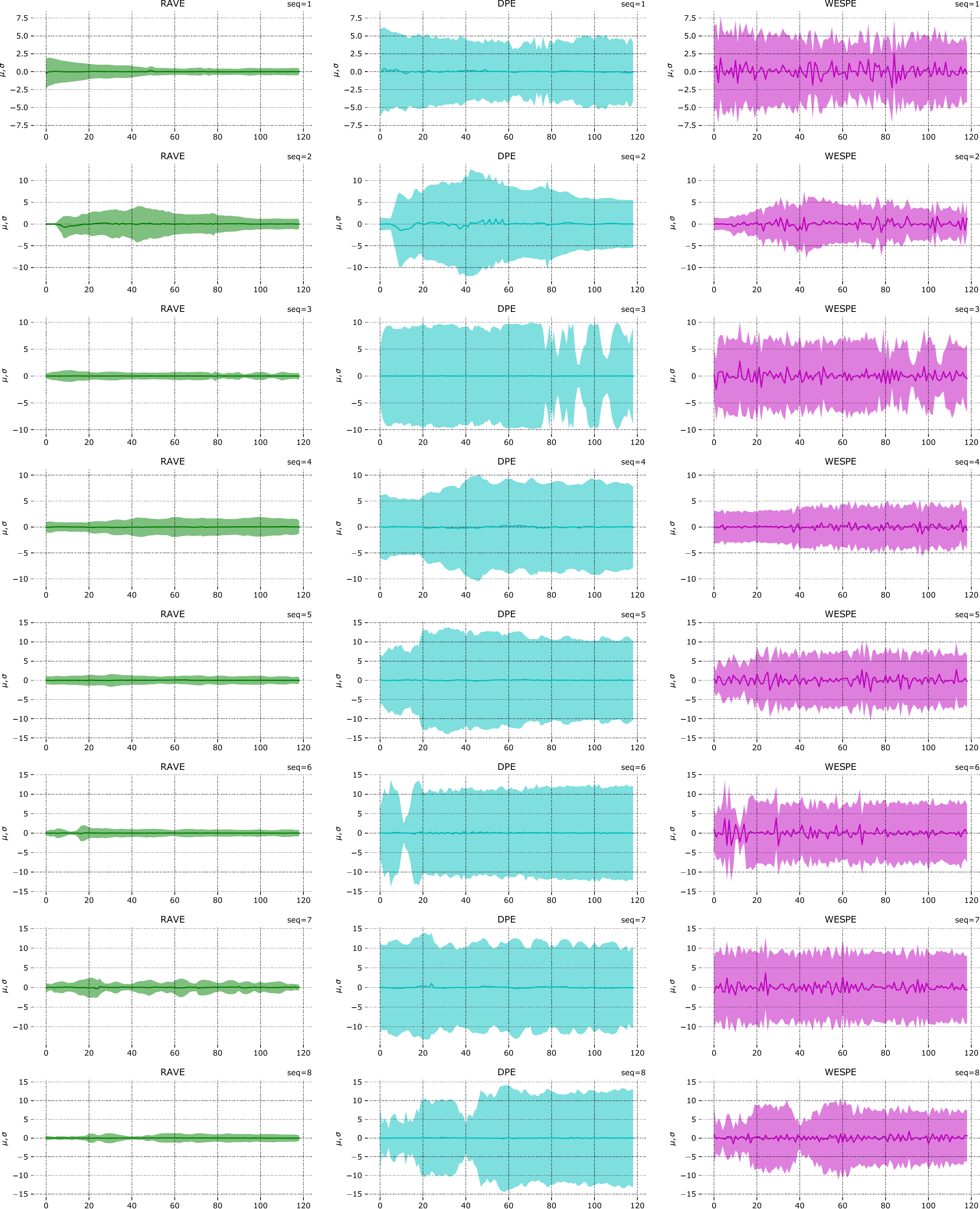}
\end{center}
   \caption{Relative pixel-flow (RPF) curves on Vid3oC dataset (Sequences 1-8). Mean $\mu_{d_t}$ (solid lines) and standard deviation $\sigma_{d_t}$ (borders of filled area) of pixel-flow differences per frame $d_t = \sum_{h}^{H} \sum_{w}^{W} F_{M}(t, h, w) - F_{S}(t, h, w)$ from method $M$ in reference to source $S$ sequences. Ripples in $\mu$ and $\sigma$ curves indicate temporal inconsistencies, the amplitude shows the extent of the artifacts. See Sec.~\ref{sec:temporal_consistency} for implementation details.}
\label{fig:rpf_vid3oc}
\end{figure*}

\begin{figure*}[ht!]
\begin{center}
\textbf{RPF Curves - Huawei}\par\medskip
\includegraphics[width=0.85\linewidth]{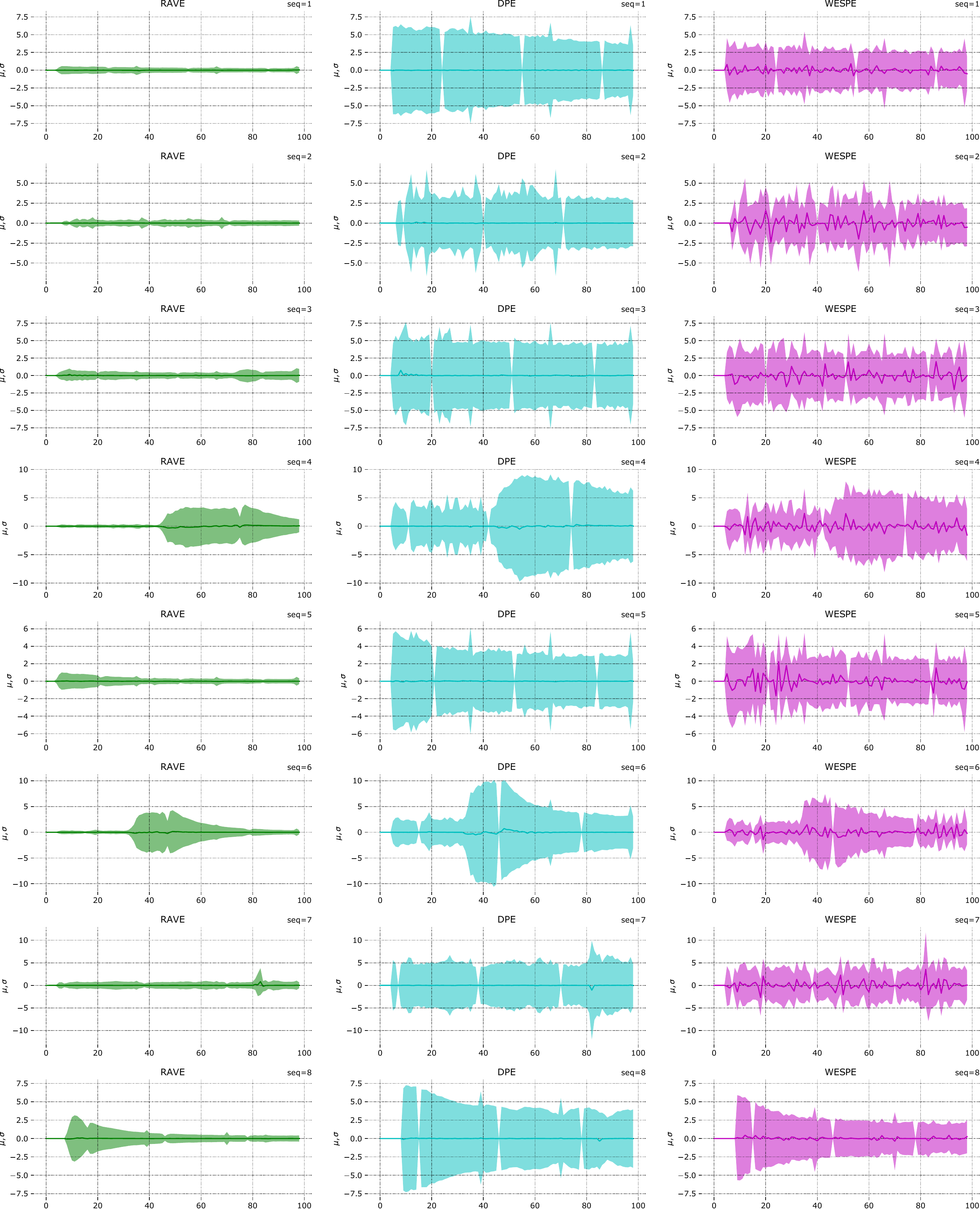}
\end{center}
   \caption{Relative pixel-flow (RPF) curves on Huawei dataset (Sequences 1-8). Mean $\mu_{d_t}$ (solid lines) and standard deviation $\sigma_{d_t}$ (borders of filled area) of pixel-flow differences per frame $d_t = \sum_{h}^{H} \sum_{w}^{W} F_{M}(t, h, w) - F_{S}(t, h, w)$ from method $M$ in reference to source $S$ sequences. Ripples in $\mu$ and $\sigma$ curves indicate temporal inconsistencies, the amplitude shows the extent of the artifacts. See Sec.~\ref{sec:temporal_consistency} for implementation details. 
   }
\label{fig:rpf_huawei}
\end{figure*}

We compare all methods by assessing the mean flow $\mu_{d_t}$ and standard deviation $\sigma_{d_t}$ of our suggested RPF in the first 8 sequences from the Vid3oC and Huawei validation sets. The curves are plotted in Figs.~\ref{fig:rpf_vid3oc},\ref{fig:rpf_huawei}, and show a clear advantage of our model against single frame enhancers DPE and WESPE. Both exhibit more serious artifacts in the temporal domain. 
RAVE generates the most consistent video, as indicated by a significantly lower standard deviation $\sigma_{d_t}$, together with a smooth evolution of $\mu_{d_t}$ and $\sigma_{d_t}$ over time.
Both DPE and WESPE expose more ripples than RAVE. Additionally, WESPE's RPF-curves show clearly higher ripples in the mean, which is indication for flickering artifacts (illumination). These temporal artifacts are most likely caused by inconsistent enhancement due to isolated single frame enhancement. In that regard, the curves show equal behavior on both datasets.

\subsection{Ablation Study}
Due to the advantage of being able to provide LPIPS scores, we evaluate 6 different architectures on Vid3oC to show the advantages of our model. All experiments are conducted with the same hyperparameter settings. We show the advantage of our modified RaGAN* objective by comparing it with the standard sigmoid cross-entropy evaluation. In combination with recurrence, RaGAN* clearly improves the enhancement results. In order to show the effectiveness of our novel LGM module, we remove all LGM modules in the generator, degrader and discriminator. Further we show the effects of recurrence to accuracy and speed in various settings by removing all recurrent connections in our networks. The results for this study are summarized in Tab.~\ref{tab:ablation}. These results show that our proposed LGM module plays a crucial part in both the recurrent and non-recurrent setting, as the configurations with LGM achieve the best scores among all configurations. This emphasizes the importance of having a global view for VQM. Our final recurrent model with LGM and RaGAN* achieves the best enhancement quality scores among all configurations.

As can be seen by comparison of runtimes in Tab.~\ref{tab:ablation}, the LGM module only introduces a marginal increase of 2 ms (non-recurrent) and 2.5ms (recurrent) relative to their non-LGM configurations respectively. This shows the effectiveness of coupled recurrence and LGM, which eventually facilitates high-quality real-time video enhancement in FullHD.

Moreover, we computed the complexity of our single discriminator in comparison to a two discriminator setting in Tab.~\ref{tab:ablation_discriminator}. The difference in FID is small in relation to the significantly larger complexity in the two discriminator setting (approximately factor $\times 2$ in all metrics). Our single discriminator largely reduces the computational demand with a comparatively small decrease in performance only.

\begin{table}[ht!]
  \caption{Comparison between single discriminator and two discriminators setting. We compare number of parameters (\#Par.), MACs (multiply-accumulate operations), FLOPs (floating-point operations), memory requirements (Mem. Req.) and FID score on Vid3oC.}
  \label{tab:ablation_discriminator}
  \centering
  \resizebox{\linewidth}{!}{
  \begin{tabular}{lrrrrr}
    \toprule
     Discriminator(s)& $\downarrow$\#Par. & $\downarrow$MACs & $\downarrow$FLOPs & $\downarrow$Mem. Req. & $\downarrow$FID \\
    \midrule
    Single& 215.12k & 71.18G & 142.60G & 1’736.88MB & 51.85 \\
    Two& 411.96k & 136.32G & 273.12G & 3’397.65MB &  51.25\\
    \bottomrule
  \end{tabular}
  }
\end{table}

We further conducted experiments to determine the impact of the ratio between local and global channels in our LGM module. We get the following FID scores for a variety of global channel sizes; 4/52.20, 8/51.85, 16/51.88, 32/51.82, 48/54.85, 64/55.31 (global channel size/FID score). Large global channel sizes (48, 64) and the smallest global channel size (4) achieve the worst FID scores, sizes in-between achieve similar scores. Our models are trained with a global channel size of 8, which provides a good balance between global and local processing with good visual quality and speed.

\section{Conclusion}
\label{sec:conclusion}
In this paper, we introduced a novel GAN-based framework, named RAVE, for unsupervised video enhancement. RAVE is designed to effectively deal with the major practical challenges associated with the video enhancement task. 
More specifically, our framework allows efficient, unsupervised, and data-driven learning from unpaired video sequences, which  does not make any impractical demand of aligned data collection for supervision.
Moreover, RAVE produces spatio-temporally consistent output thanks to the  enhanced guidance of the proposed recurrent architecture and LGM module.
Efficiency in memory and computation (both during  training and inference) was achieved by  means of the joint-distribution learning strategy and our computationally efficient recurrent cells.

Consequently, our data-driven method performs the quality mapping without any strong inductive bias, which is also reflected by its training stability with no need to adjust the hyperparameter $\alpha$ across datasets. Our experiments on two datasets demonstrate that RAVE produces consistently better output than the compared methods, both in terms of  quantitative and qualitative measures.  In fact, RAVE  is the first method that achieves real-time performance while addressing the problem of  temporal consistency. In addition to the mentioned benefits, the proposed framework is also generic by design. Therefore, RAVE might also be adapted to other video enhancement/manipulation tasks, merely by replacing the suitable cells.

\noindent\textbf{Acknowledgements.}
This work was partly supported by a Huawei project, the ETH Z\"urich General Fund (OK) and the Alexander von Humboldt Foundation.

{\small
\bibliographystyle{name}
\bibliography{ref}
}

\newpage
\appendix

\section*{\Large Supplementary Material}

We provide additional results, that did not fit in the main paper. We include large versions of the visual results provided in the paper, a comparison to a video translation method, present more relative pixel-flow (RPF) curves to complete the full set from both datasets and show examples of artifacts.



\section{Artifacts}

\begin{figure*}[t]
\begin{center}
\includegraphics[width=\linewidth]{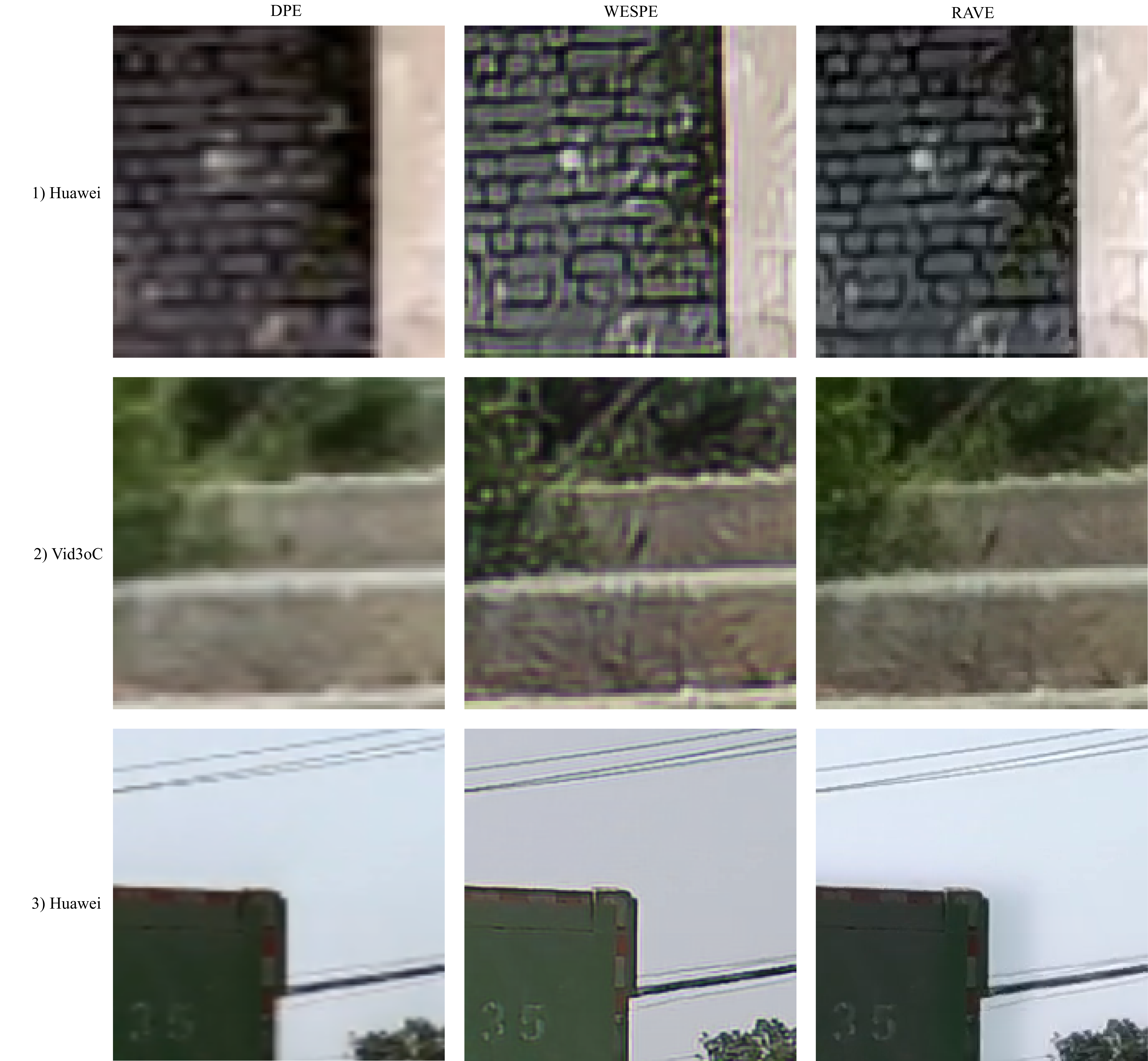}
\end{center}
   \caption{Artifacts comparison. Artifacts shown in the first and second row for DPE and WESPE are present in all frames on both datasets. The third row shows an isolated case of artifacts for RAVE (limited to large movements on Huawei).
   }
\label{fig:artifacts}
\end{figure*}

As discussed in the paper, both DPE and WESPE show severe artifacts on both datasets on frame-level as well as in the temporal domain. We show some close-ups from examples on both datasets in Fig.~\ref{fig:artifacts}. As shown by the RPF curves, RAVE is not only much more resilient with regard to artifacts in the temporal domain, but also on frame-level. We did not observe any serious artifacts on Vid3oC for RAVE. There are some artifacts in specific cases, that are related to very large movements on the Huawei dataset. These cases however, are limited to areas with large movement together with abrupt changes of color (large steps in the signal), see third row in Fig.~\ref{fig:artifacts}. Since our proposed method (RAVE) is data-driven and these effects are not present on Vid3oC -- which generally contains more movement than Huawei -- we are convinced this can be easily resolved by collecting more fast motion data for the Huawei dataset.
DPE produces severely blurry frames due to its fixed input resolution, while WESPE exposes severe low-level artifacts, see first and second row in Fig.~\ref{fig:artifacts}. There are green patterns distributed over all frames in both datasets. Also, there are ringing artifacts around edges as can be seen in the third row.

\section{RecycleGAN}

\begin{figure*}[h!]
\begin{center}
\includegraphics[width=\linewidth]{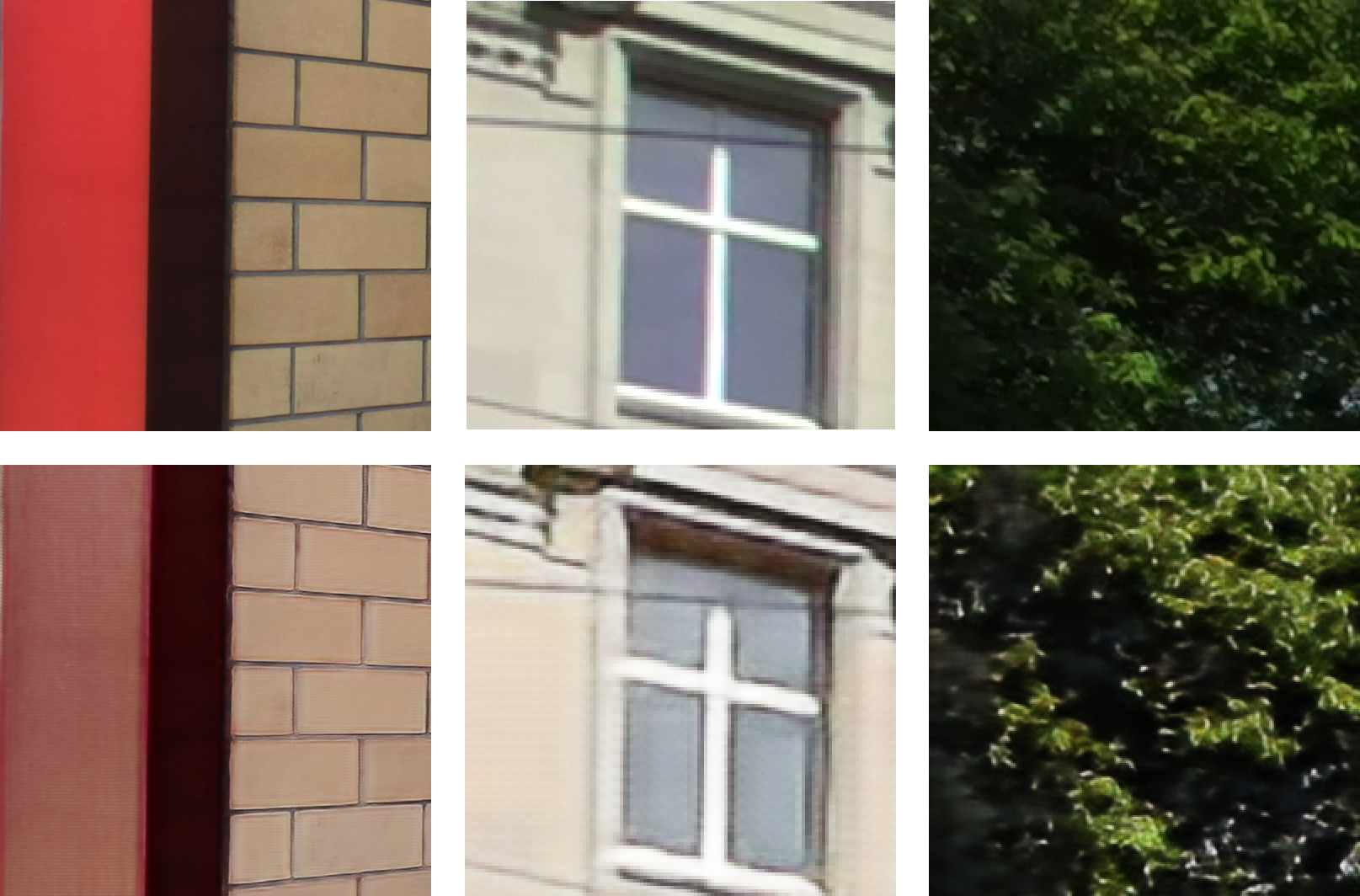}
\end{center}
 \caption{Examples of RAVE (top row) and RecycleGAN (bottom row) on the Vid3oC dataset. The examples show the clear benefit of our method. RecycleGAN exposes unpleasant artifacts, distortions and color shifts, which are not present in our method.}
\label{fig:recyclegan}
\end{figure*}

To the best of our knowledge, we are the first to design a video-based method for video quality mapping. In order to better understand the advantages of our proposed method for high-resolution video quality mapping, we additionally compare our model to RecycleGAN~\cite{Recycle-GAN}. The method is designed for high-level video translation tasks, which share similarities to video quality mapping. Usually, these high-level translation tasks deal with larger changes between the domains compared to more detailed enhancement in the case of image/video enhancement with subtle differences between the domains. We modified and adapted \cite{Recycle-GAN} to facilitate high-resolution video and trained it on Vid3oC, where it achieves an FID score of 56.68. This score is significantly worse than what our model attains (51.85) and also the visual quality is unsatisfactory due to unpleasant artifacts, distortions and color shifts, see Fig.~\ref{fig:recyclegan}. The experiment shows, that our video enhancement task differs from typical translation tasks and highlights the need for a specialized solution, which we provide with our proposed framework.

\section{RPF curves}
\begin{figure*}[t]
\begin{center}
\textbf{RPF Curves - Vid3oC}\par\medskip
\includegraphics[width=\linewidth]{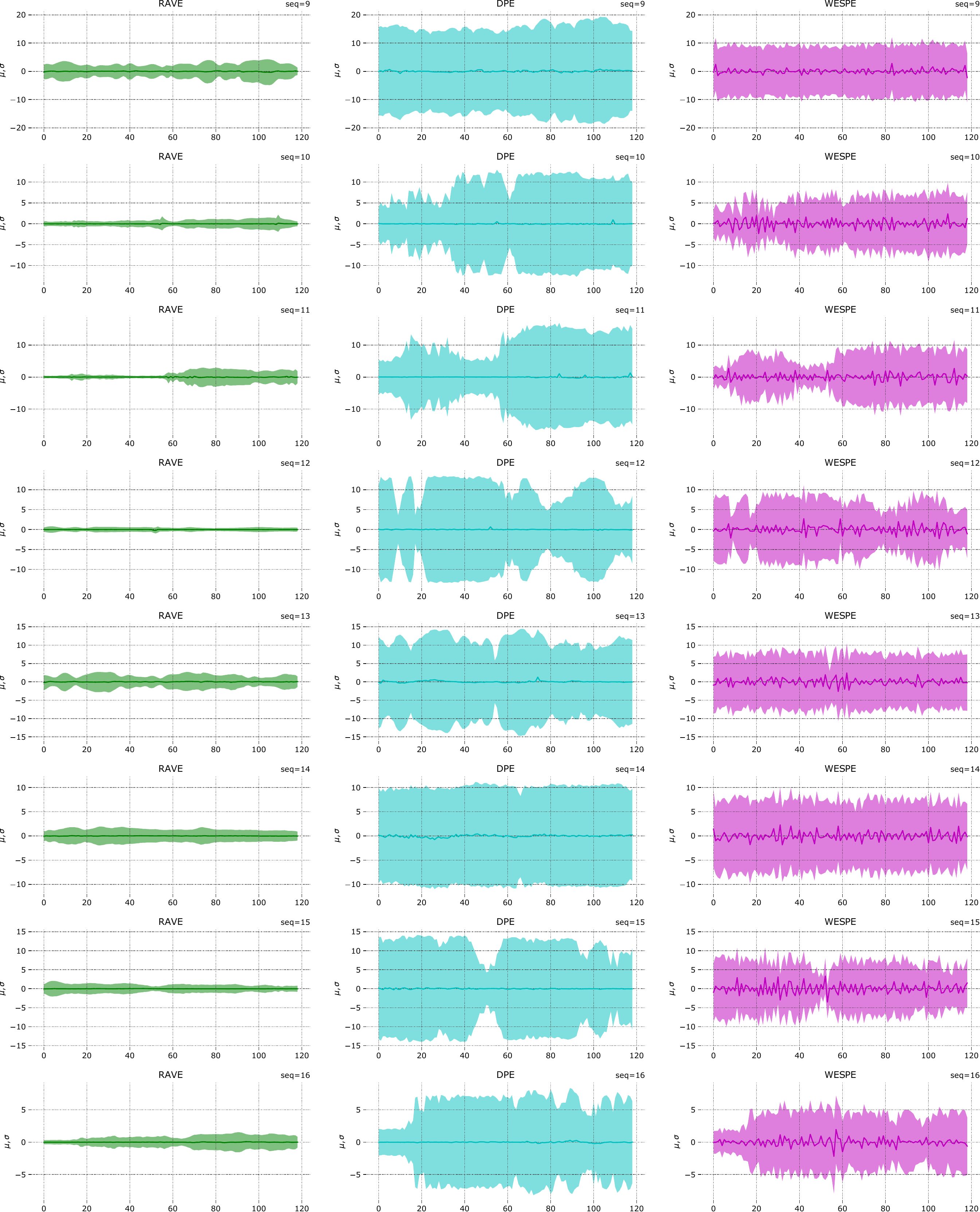}
\end{center}
   \caption{Relative pixel-flow (RPF) curves on Vid3oC dataset (Sequences 9-16). Mean $\mu_{d_t}$ (solid lines) and standard deviation $\sigma_{d_t}$ (borders of filled area) of pixel-flow differences per frame $d_t = \sum_{h}^{H} \sum_{w}^{W} F_{M}(t, h, w) - F_{S}(t, h, w)$ from method $M$ in reference to source $S$ sequences. Ripples in $\mu$ and $\sigma$ curves indicate temporal inconsistencies, the amplitude shows the extent of the artifacts.}
\label{fig:rpf_vid3oc_supp}
\end{figure*}

\begin{figure*}[t]
\begin{center}
\textbf{RPF Curves - Huawei}\par\medskip
\includegraphics[width=\linewidth]{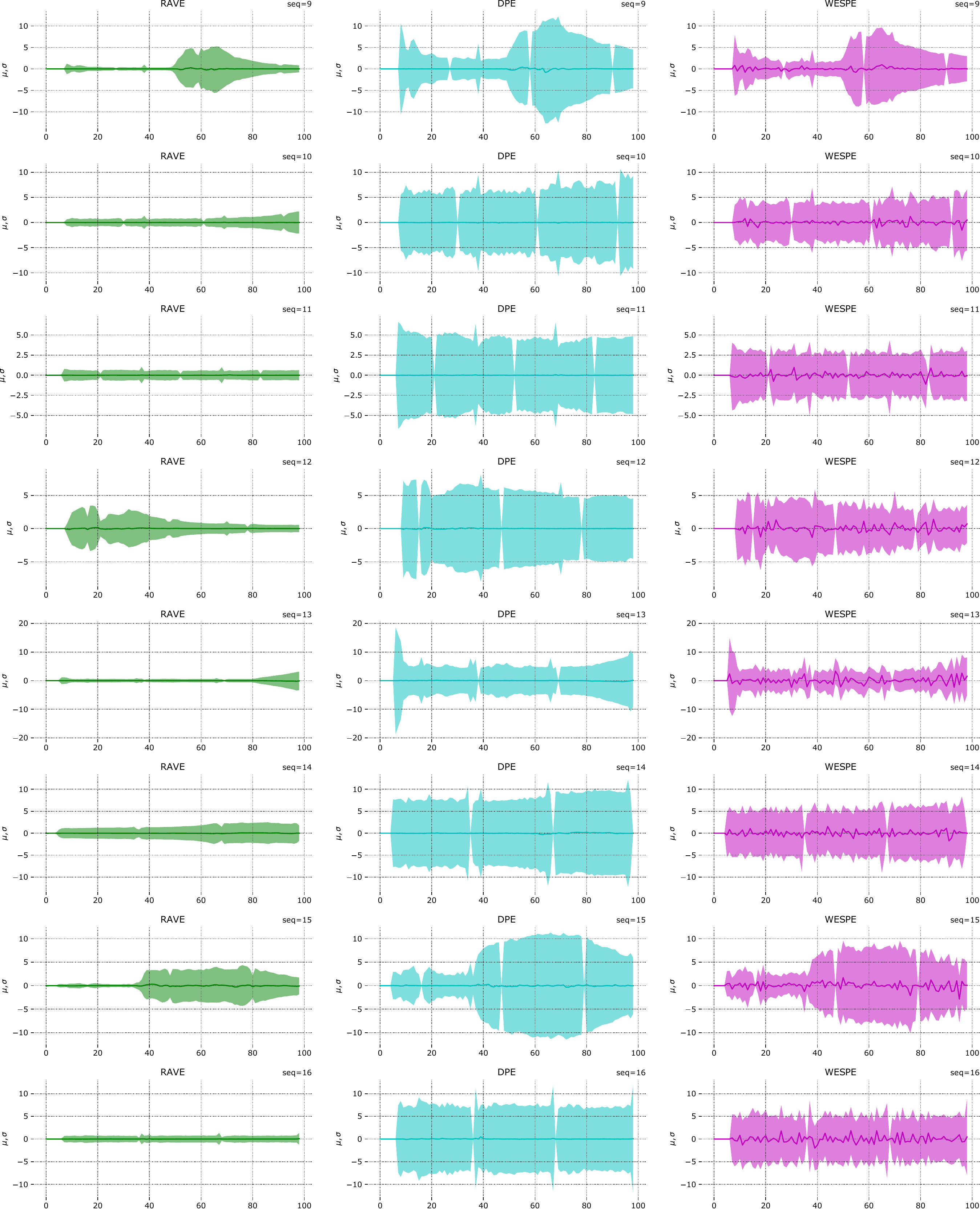}
\end{center}
   \caption{Relative pixel-flow (RPF) curves on Huawei dataset (Sequences 9-16). Mean $\mu_{d_t}$ (solid lines) and standard deviation $\sigma_{d_t}$ (borders of filled area) of pixel-flow differences per frame $d_t = \sum_{h}^{H} \sum_{w}^{W} F_{M}(t, h, w) - F_{S}(t, h, w)$ from method $M$ in reference to source $S$ sequences. Ripples in $\mu$ and $\sigma$ curves indicate temporal inconsistencies, the amplitude shows the extent of the artifacts.
   }
\label{fig:rpf_huawei_supp}
\end{figure*}

To complete the full set of RPF curves, we provide curves from videos 9-16 in Fig.~\ref{fig:rpf_vid3oc_supp} (Vid3oC) and Fig.~\ref{fig:rpf_huawei_supp} (Huawei). The curves show similar performance to the curves shown in the paper; RAVE clearly produces the temporally most consistent output compared to both other methods DPE and WESPE.

\section{Large Visual Examples on Vid3oC and Huawei}

\begin{figure*}[t]
\begin{center}
\includegraphics[width=\linewidth]{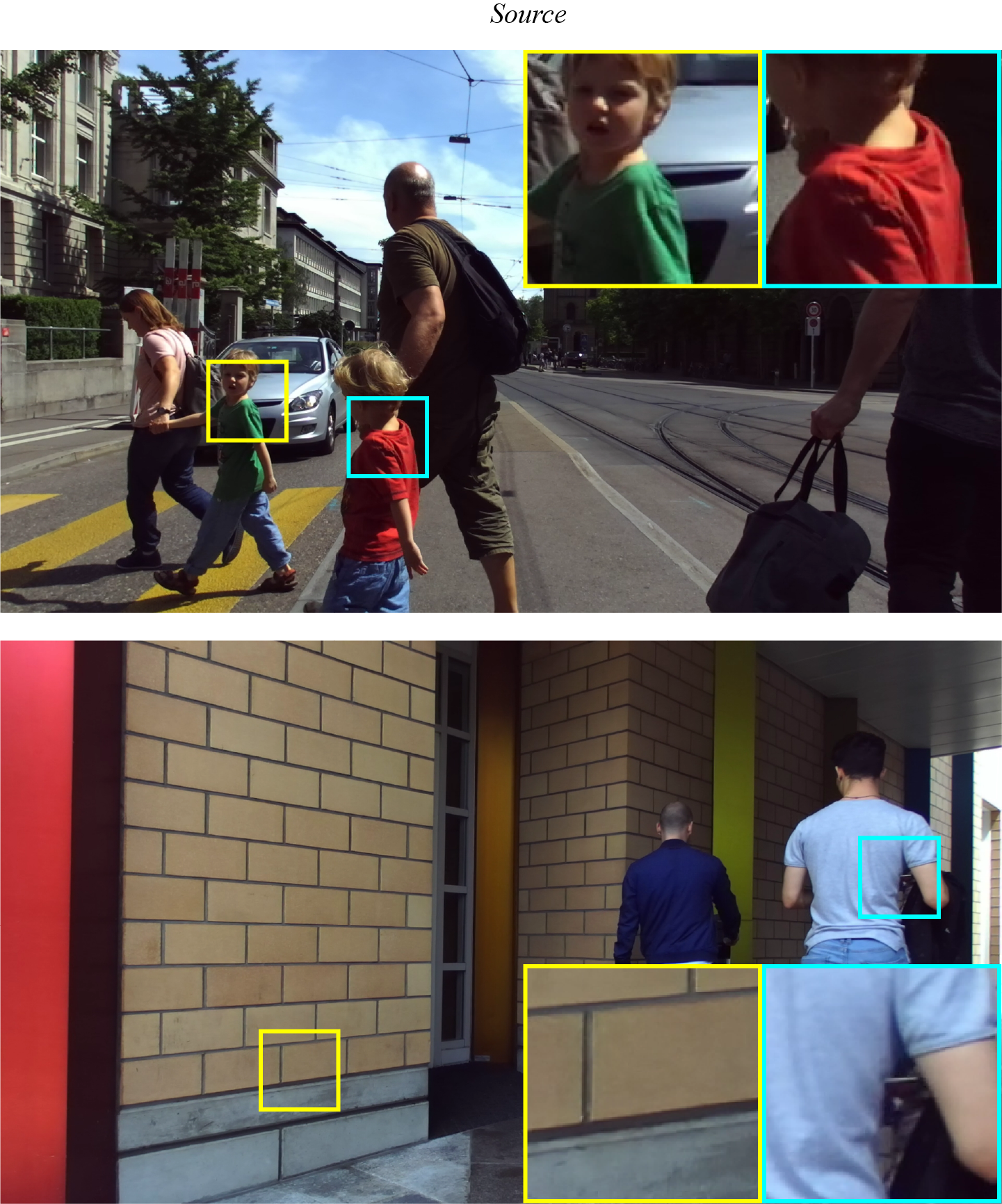}
\end{center}
 \caption{Vid3oC - Source Examples.}
\label{fig:vid3oc_start}
\end{figure*}

\begin{figure*}[t]
\begin{center}
\includegraphics[width=\linewidth]{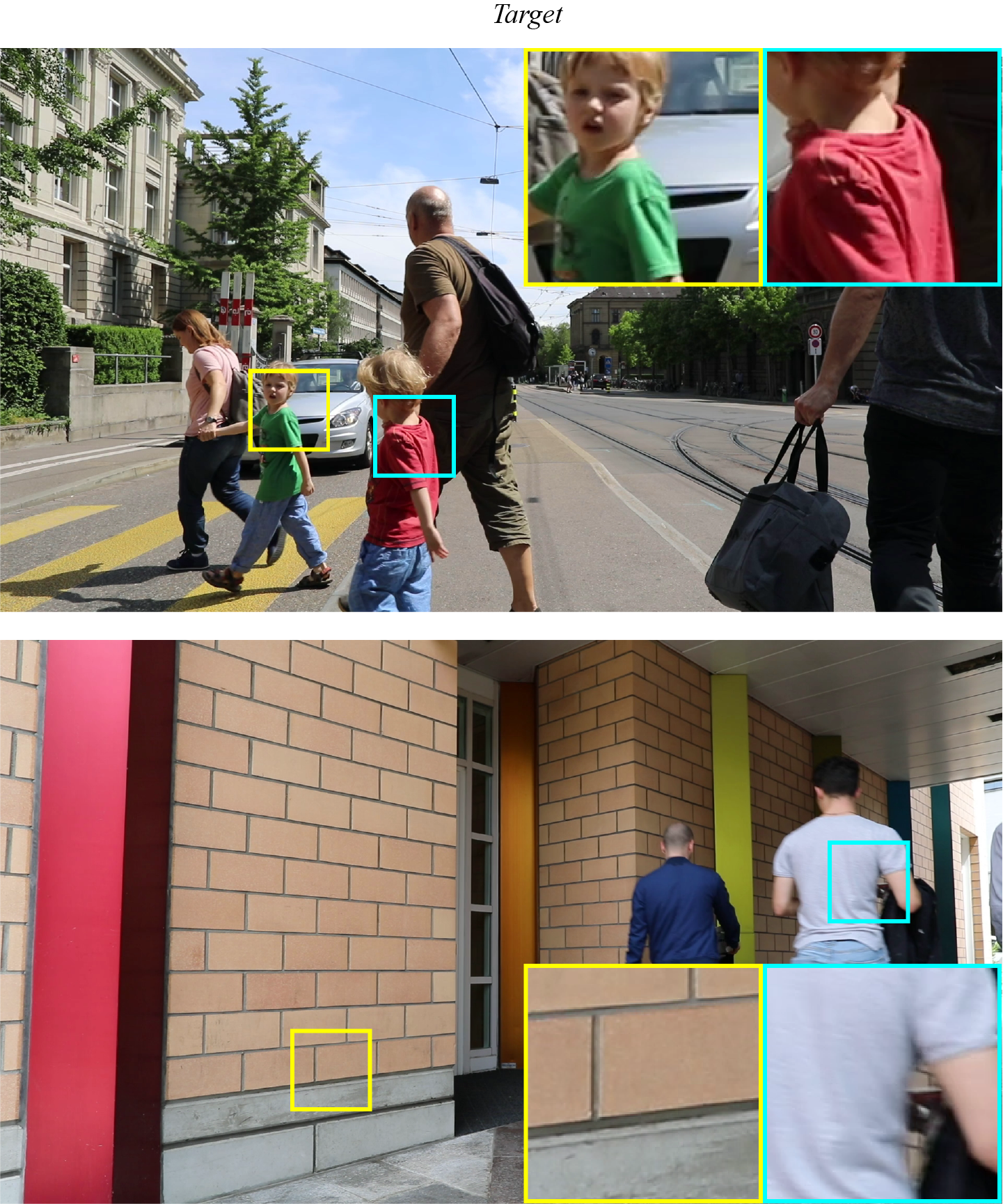}
\end{center}
 \caption{Vid3oC - Target Examples.}
\label{fig:vid3oc_end}
\end{figure*}

\begin{figure*}[t]
\begin{center}
\includegraphics[width=\linewidth]{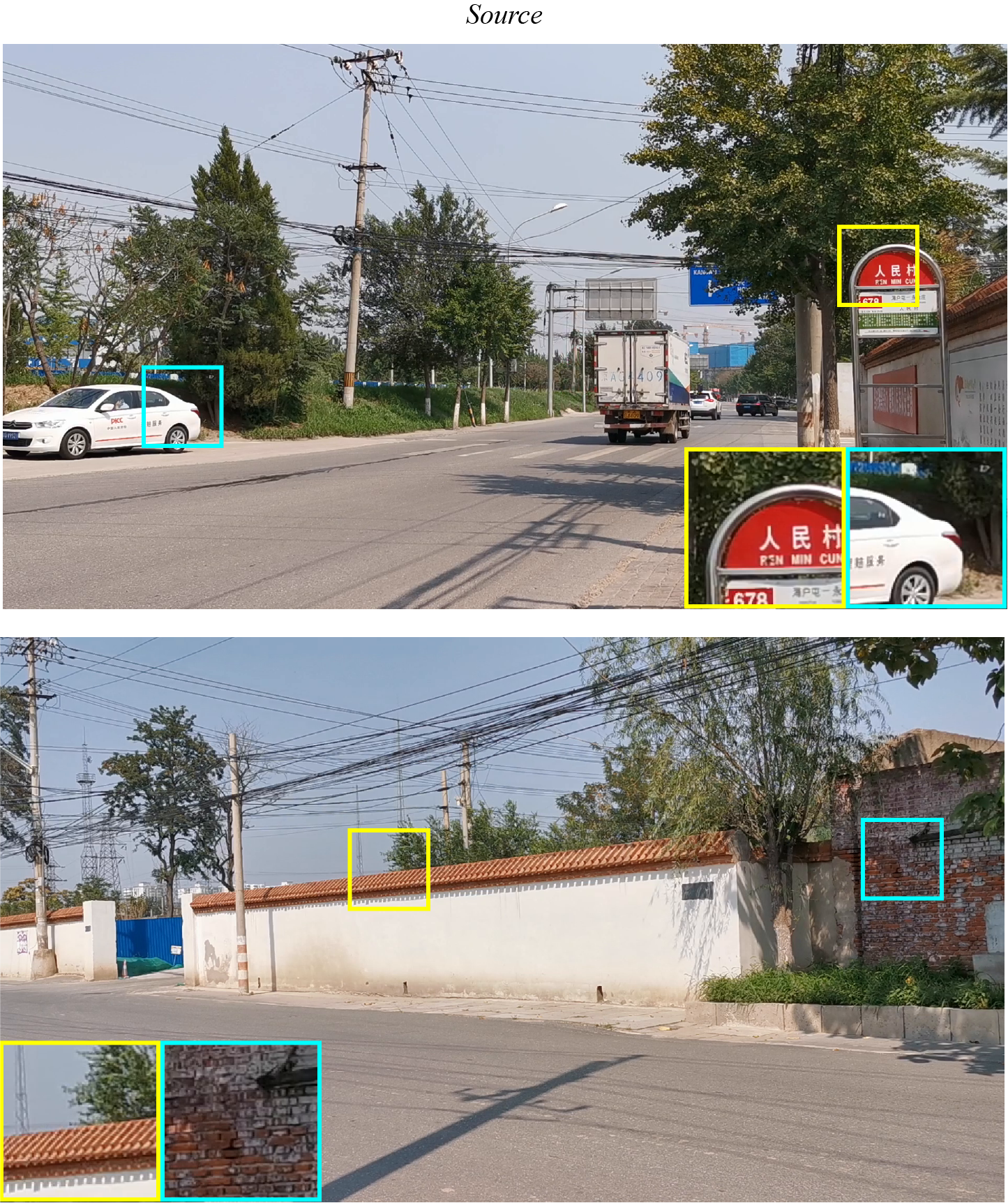}
\end{center}
 \caption{Huawei - Source Examples.}
\label{fig:huawei_start}
\end{figure*}

\begin{figure*}[t]
\begin{center}
\includegraphics[width=\linewidth]{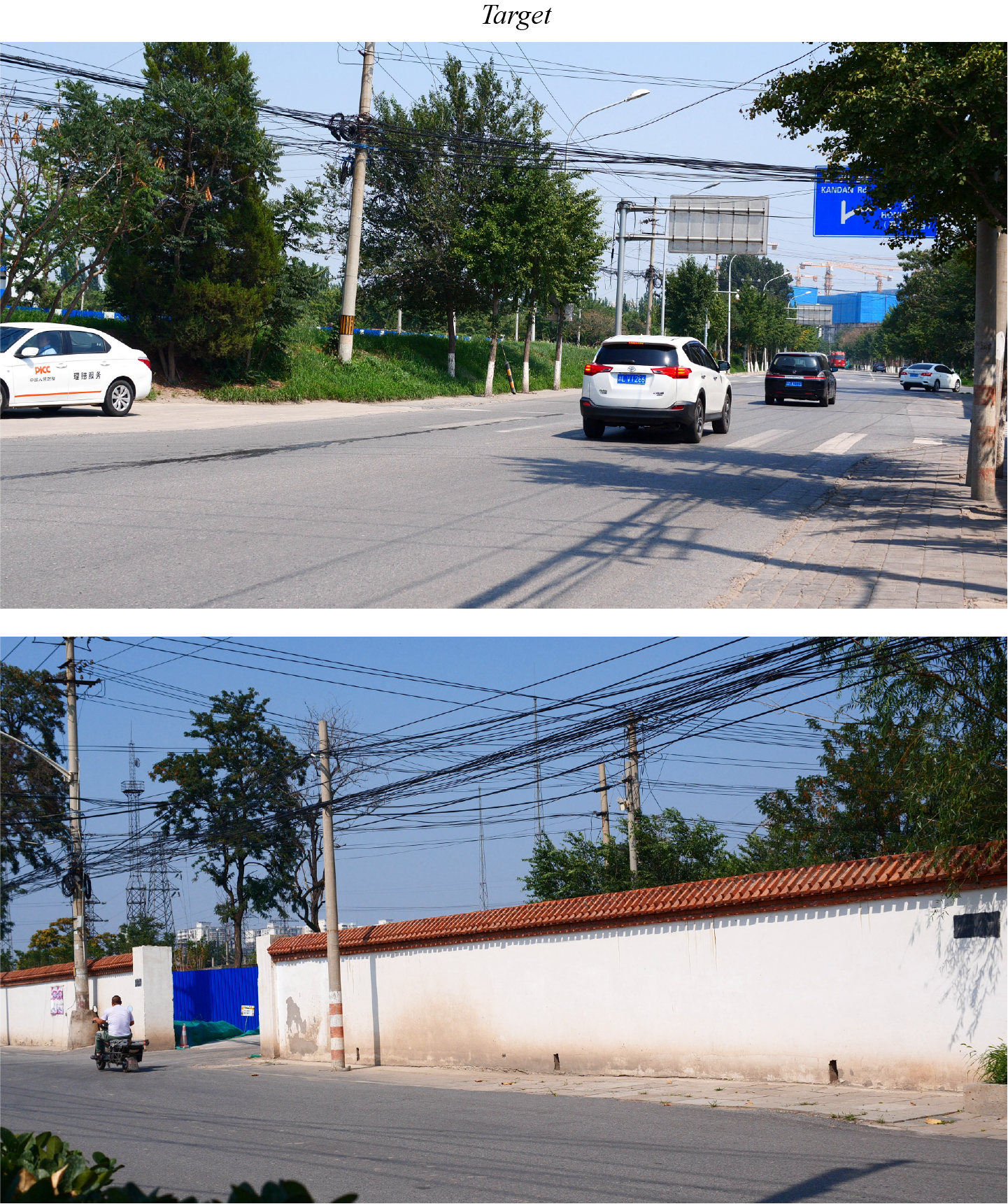}
\end{center}
 \caption{Huawei - Target Examples.}
\label{fig:huawei_end}
\end{figure*}

We provide larger versions of our visual examples from the main paper for Source, WESPE~\cite{ignatov2018wespe}, DPE~\cite{chen2018deep}, RAVE and Target in Figs.~\ref{fig:vid3oc_start}-\ref{fig:huawei_end} from both Vid3oC~\cite{kim2019vid3oc} and Huawei dataset.

\begin{figure*}[t]
\begin{center}
\includegraphics[width=\linewidth]{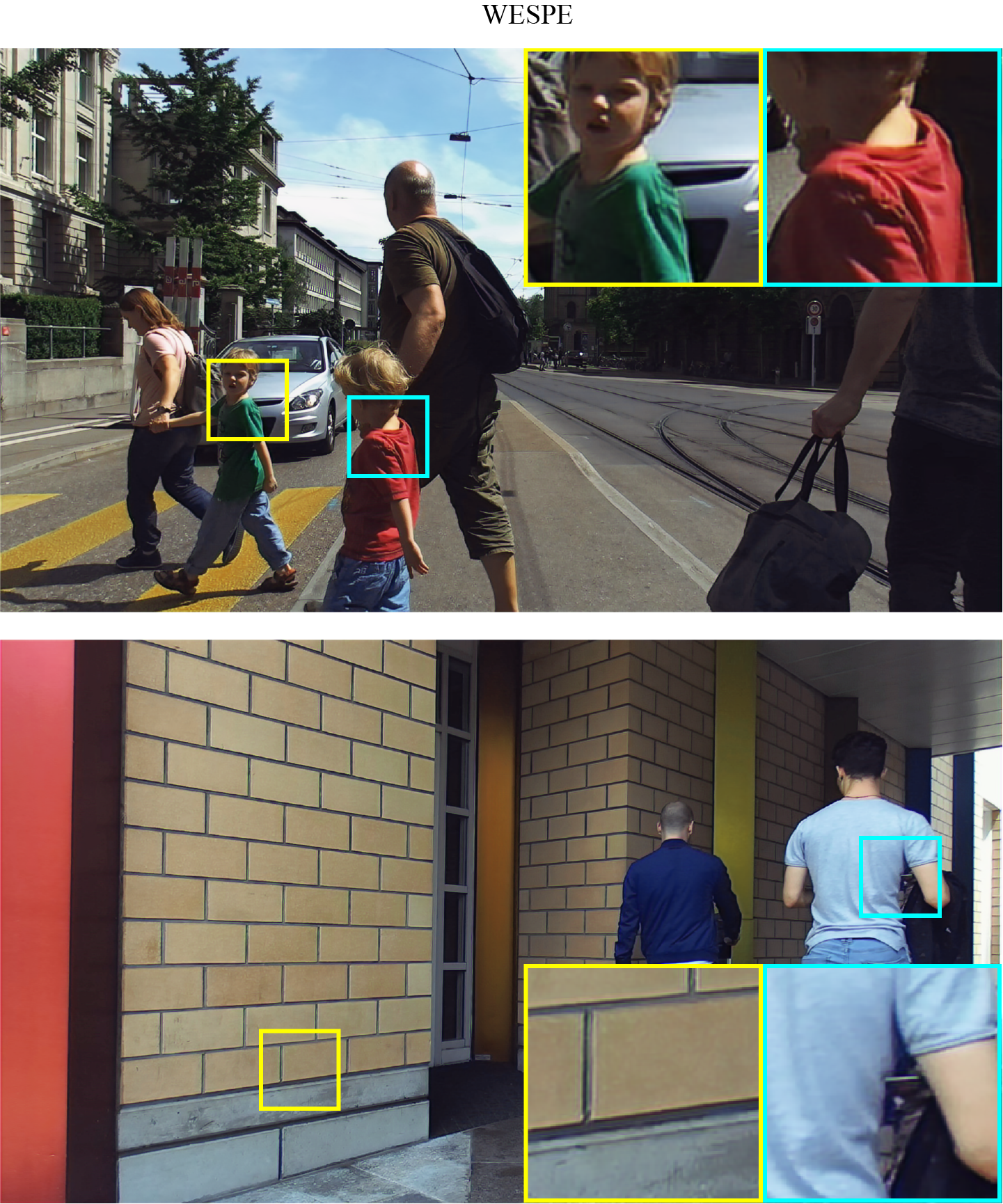}
\end{center}
 \caption{Vid3oC - WESPE Examples.}
\end{figure*}

\begin{figure*}[t]
\begin{center}
\includegraphics[width=\linewidth]{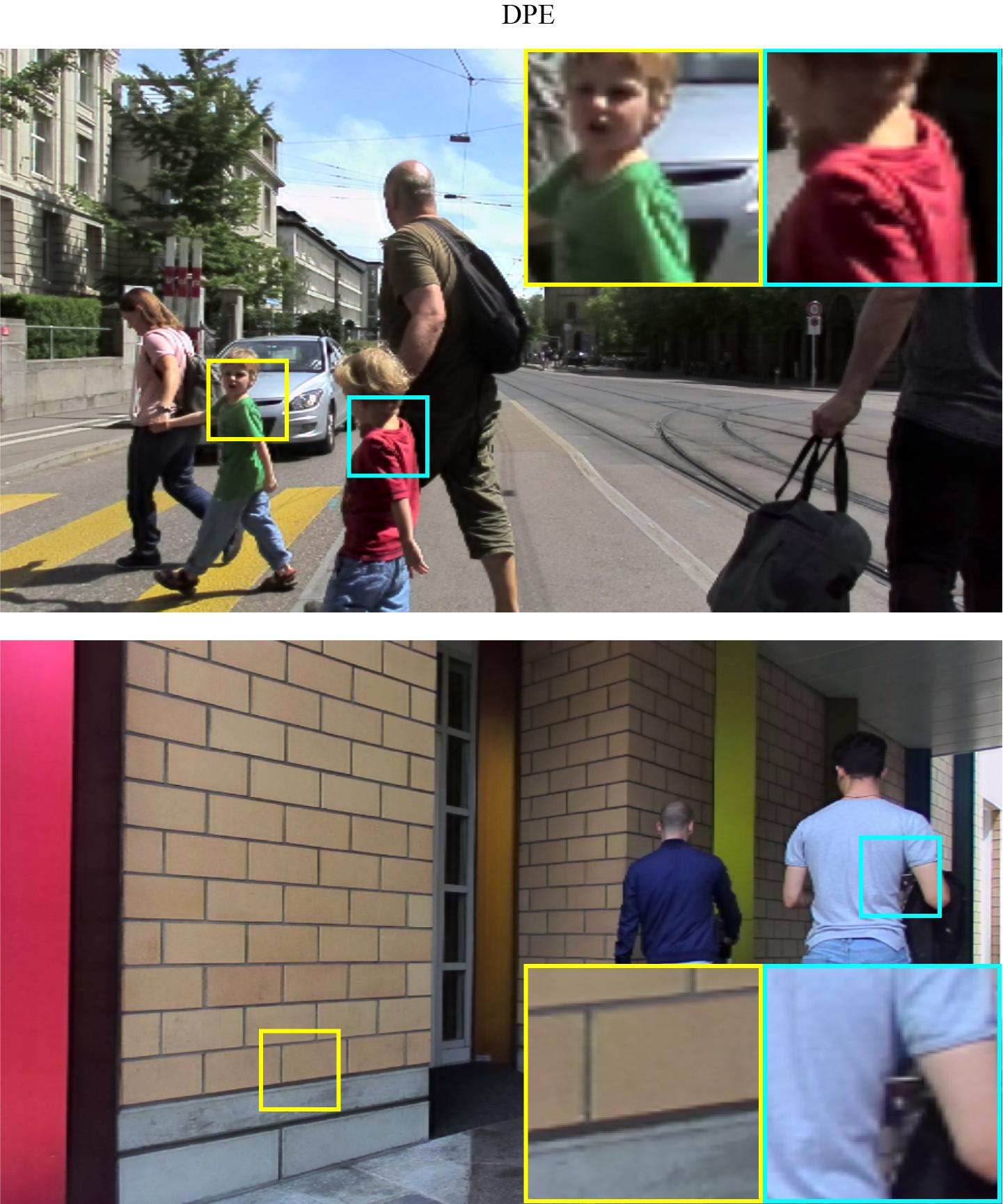}
\end{center}
 \caption{Vid3oC - DPE Examples.}
\end{figure*}

\begin{figure*}[t]
\begin{center}
\includegraphics[width=\linewidth]{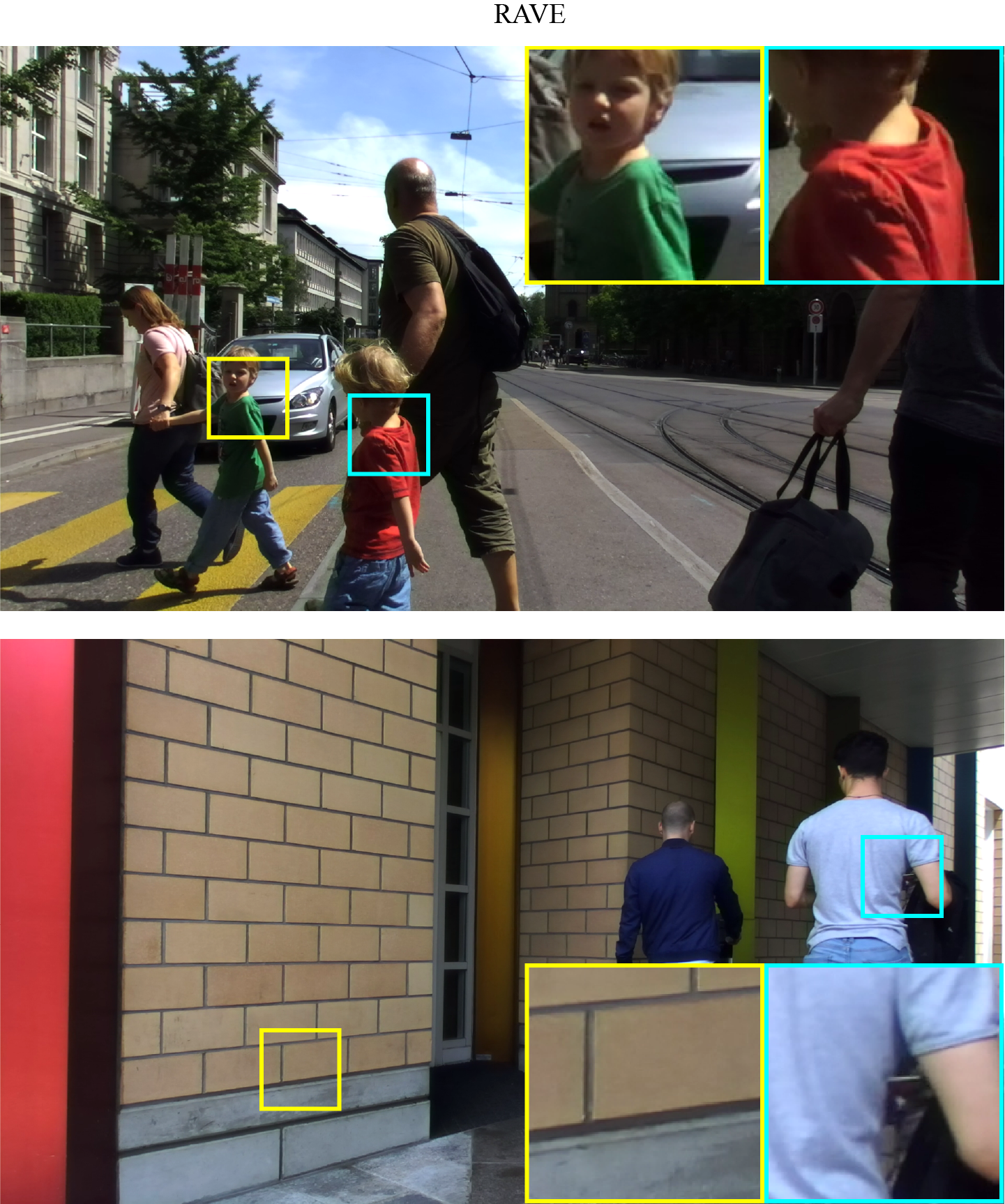}
\end{center}
 \caption{Vid3oC - RAVE Examples.}
\end{figure*}

\begin{figure*}[t]
\begin{center}
\includegraphics[width=\linewidth]{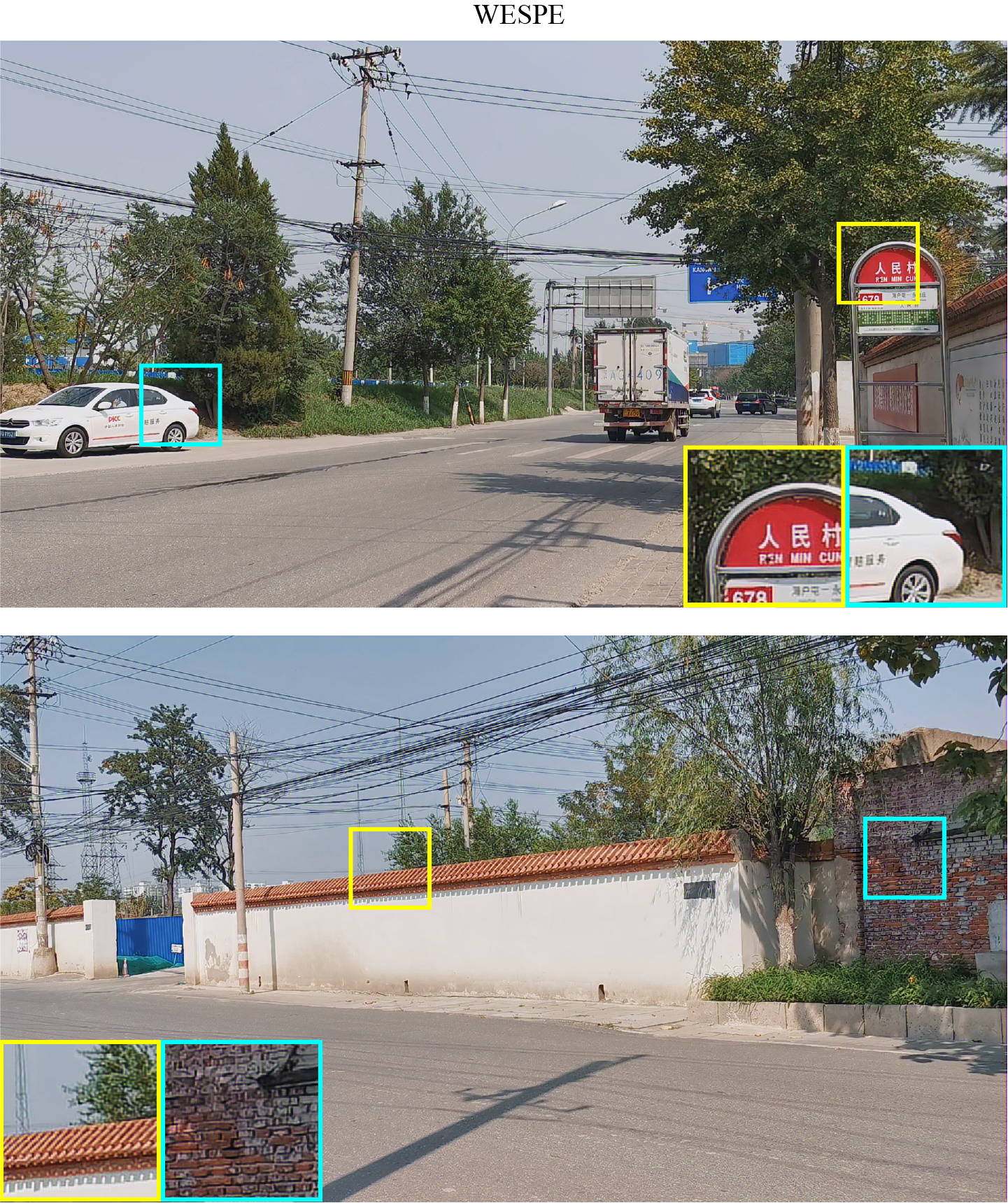}
\end{center}
 \caption{Huawei - WESPE Examples.}
\end{figure*}

\begin{figure*}[t]
\begin{center}
\includegraphics[width=\linewidth]{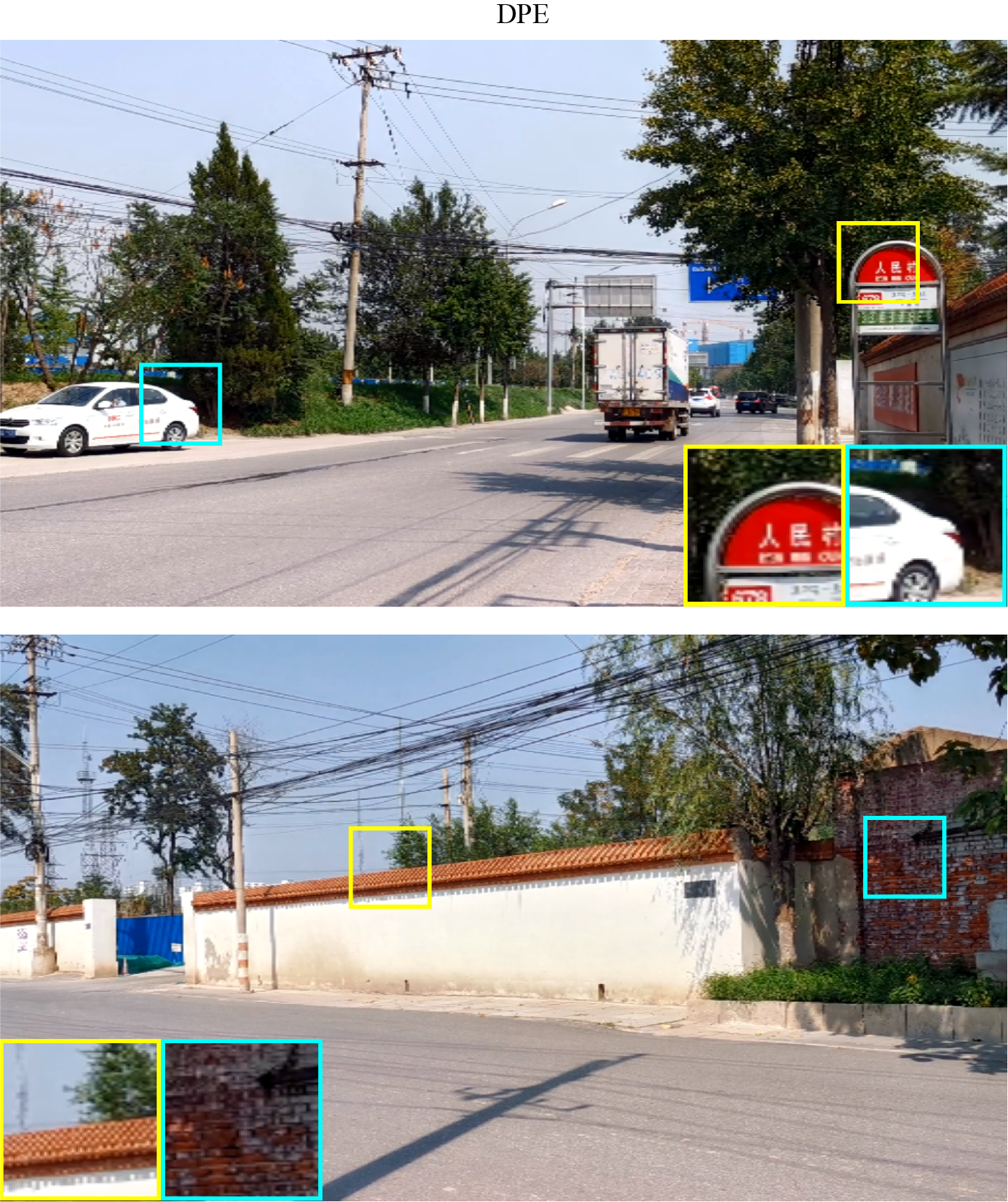}
\end{center}
 \caption{Huawei - DPE Examples.}
\end{figure*}

\begin{figure*}[t]
\begin{center}
\includegraphics[width=\linewidth]{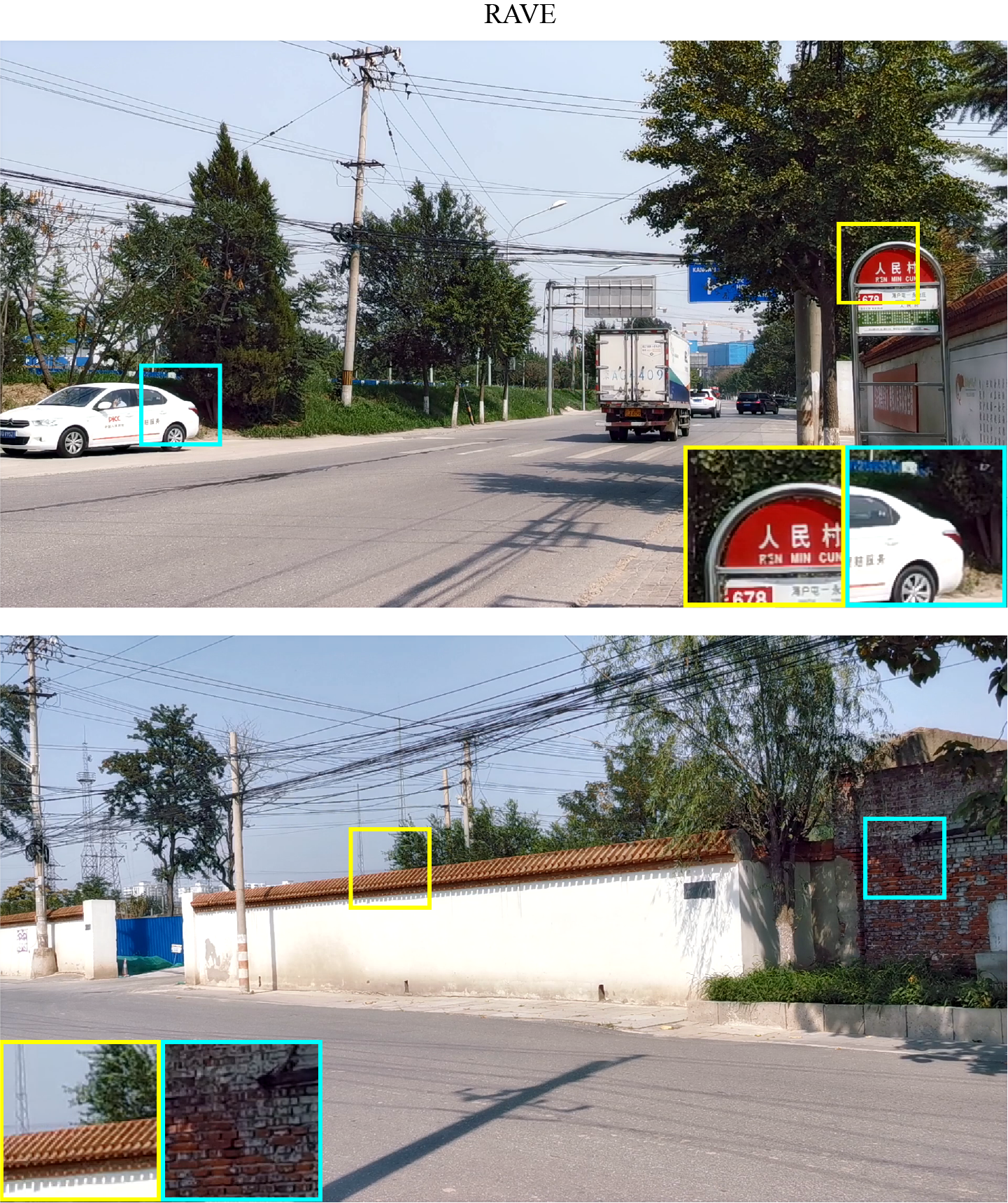}
\end{center}
 \caption{Huawei - RAVE Examples.}
\end{figure*}

\end{document}